



\font\bigbold=cmbx12
\font\ninerm=cmr9
\font\eightrm=cmr8
\font\sixrm=cmr6
\font\fiverm=cmr5
\font\ninebf=cmbx9
\font\eightbf=cmbx8
\font\sixbf=cmbx6
\font\fivebf=cmbx5
\font\ninei=cmmi9  \skewchar\ninei='177
\font\eighti=cmmi8  \skewchar\eighti='177
\font\sixi=cmmi6    \skewchar\sixi='177
\font\fivei=cmmi5
\font\ninesy=cmsy9 \skewchar\ninesy='60
\font\eightsy=cmsy8 \skewchar\eightsy='60
\font\sixsy=cmsy6   \skewchar\sixsy='60
\font\fivesy=cmsy5
\font\nineit=cmti9
\font\eightit=cmti8
\font\ninesl=cmsl9
\font\eightsl=cmsl8
\font\ninett=cmtt9
\font\eighttt=cmtt8
\font\tenfrak=eufm10
\font\ninefrak=eufm9
\font\eightfrak=eufm8
\font\sevenfrak=eufm7
\font\fivefrak=eufm5
\font\tenbb=msbm10
\font\ninebb=msbm9
\font\eightbb=msbm8
\font\sevenbb=msbm7
\font\fivebb=msbm5
\font\tensmc=cmcsc10


\newfam\bbfam
\textfont\bbfam=\tenbb
\scriptfont\bbfam=\sevenbb
\scriptscriptfont\bbfam=\fivebb
\def\Bbb{\fam\bbfam}

\newfam\frakfam
\textfont\frakfam=\tenfrak
\scriptfont\frakfam=\sevenfrak
\scriptscriptfont\frakfam=\fivefrak
\def\frak{\fam\frakfam}

\def\smc{\tensmc}


\def\eightpoint{%
\textfont0=\eightrm   \scriptfont0=\sixrm
\scriptscriptfont0=\fiverm  \def\rm{\fam0\eightrm}%
\textfont1=\eighti   \scriptfont1=\sixi
\scriptscriptfont1=\fivei  \def\oldstyle{\fam1\eighti}%
\textfont2=\eightsy   \scriptfont2=\sixsy
\scriptscriptfont2=\fivesy
\textfont\itfam=\eightit  \def\it{\fam\itfam\eightit}%
\textfont\slfam=\eightsl  \def\sl{\fam\slfam\eightsl}%
\textfont\ttfam=\eighttt  \def\tt{\fam\ttfam\eighttt}%
\textfont\frakfam=\eightfrak \def\frak{\fam\frakfam\eightfrak}%
\textfont\bbfam=\eightbb  \def\Bbb{\fam\bbfam\eightbb}%
\textfont\bffam=\eightbf   \scriptfont\bffam=\sixbf
\scriptscriptfont\bffam=\fivebf  \def\bf{\fam\bffam\eightbf}%
\abovedisplayskip=9pt plus 2pt minus 6pt
\belowdisplayskip=\abovedisplayskip
\abovedisplayshortskip=0pt plus 2pt
\belowdisplayshortskip=5pt plus2pt minus 3pt
\smallskipamount=2pt plus 1pt minus 1pt
\medskipamount=4pt plus 2pt minus 2pt
\bigskipamount=9pt plus4pt minus 4pt
\setbox\strutbox=\hbox{\vrule height 7pt depth 2pt width 0pt}%
\normalbaselineskip=9pt \normalbaselines
\rm}


\def\ninepoint{%
\textfont0=\ninerm   \scriptfont0=\sixrm
\scriptscriptfont0=\fiverm  \def\rm{\fam0\ninerm}%
\textfont1=\ninei   \scriptfont1=\sixi
\scriptscriptfont1=\fivei  \def\oldstyle{\fam1\ninei}%
\textfont2=\ninesy   \scriptfont2=\sixsy
\scriptscriptfont2=\fivesy
\textfont\itfam=\nineit  \def\it{\fam\itfam\nineit}%
\textfont\slfam=\ninesl  \def\sl{\fam\slfam\ninesl}%
\textfont\ttfam=\ninett  \def\tt{\fam\ttfam\ninett}%
\textfont\frakfam=\ninefrak \def\frak{\fam\frakfam\ninefrak}%
\textfont\bbfam=\ninebb  \def\Bbb{\fam\bbfam\ninebb}%
\textfont\bffam=\ninebf   \scriptfont\bffam=\sixbf
\scriptscriptfont\bffam=\fivebf  \def\bf{\fam\bffam\ninebf}%
\abovedisplayskip=10pt plus 2pt minus 6pt
\belowdisplayskip=\abovedisplayskip
\abovedisplayshortskip=0pt plus 2pt
\belowdisplayshortskip=5pt plus2pt minus 3pt
\smallskipamount=2pt plus 1pt minus 1pt
\medskipamount=4pt plus 2pt minus 2pt
\bigskipamount=10pt plus4pt minus 4pt
\setbox\strutbox=\hbox{\vrule height 7pt depth 2pt width 0pt}%
\normalbaselineskip=10pt \normalbaselines
\rm}


\def\pagewidth#1{\hsize= #1}
\def\pageheight#1{\vsize= #1}
\def\hcorrection#1{\advance\hoffset by #1}
\def\vcorrection#1{\advance\voffset by #1}

\newif\iftitlepage   \titlepagetrue               
\newtoks\titlepagefoot     \titlepagefoot={\hfil} 
\newtoks\otherpagesfoot    \otherpagesfoot={\hfil\tenrm\folio\hfil}
\footline={\iftitlepage\the\titlepagefoot\global\titlepagefalse
           \else\the\otherpagesfoot\fi}

\font\extra=cmss10 scaled \magstep0
\setbox1 = \hbox{{{\extra R}}}
\setbox2 = \hbox{{{\extra I}}}
\setbox3 = \hbox{{{\extra C}}}
\setbox4 = \hbox{{{\extra Z}}}
\setbox5 = \hbox{{{\extra N}}}

\def\RRR{{{\extra R}}\hskip-\wd1\hskip2.0 
   true pt{{\extra I}}\hskip-\wd2
\hskip-2.0 true pt\hskip\wd1}
\def\Real{\hbox{{\extra\RRR}}}    

\def\CCC{{{\extra C}}\hskip-\wd3\hskip 2.5 true pt{{\extra I}}
\hskip-\wd2\hskip-2.5 true pt\hskip\wd3}
\def\Complex{\hbox{{\extra\CCC}}\!\!}   

\def\ZZZ{{{\extra Z}}\hskip-\wd4\hskip 2.5 true pt{{\extra Z}}}
\def\Zed{\hbox{{\extra\ZZZ}}}       




\def\R{{\Real}}
\def\C{{\Complex}}

\def\frac#1#2{{#1\over#2}}

\def\({\left(}
\def\){\right)}
\def\<{\langle}
\def\>{\rangle}

\def\pmb#1{\setbox0=\hbox{$#1$}%
   \kern-.025em\copy0\kern-\wd0
   \kern.05em\copy0\kern-\wd0
   \kern-.025em\raise.0433em\box0 }


\def\abstract#1{{\parindent=30pt\narrower\noindent\ninepoint\openup
2pt #1\par}}


\newcount\notenumber\notenumber=1
\def\note#1
{\unskip\footnote{$^{\the\notenumber}$}
{\eightpoint\openup 1pt #1}
\global\advance\notenumber by 1}


\global\newcount\secno \global\secno=0
\global\newcount\meqno \global\meqno=1
\global\newcount\appno \global\appno=0
\newwrite\eqmac
\def\romappno{\ifcase\appno\or A\or B\or C\or D\or E\or F\or G\or H
\or I\or J\or K\or L\or M\or N\or O\or P\or Q\or R\or S\or T\or U\or
V\or W\or X\or Y\or Z\fi}
\def\eqn#1{
        \ifnum\secno>0
            \eqno(\the\secno.\the\meqno)\xdef#1{\the\secno.\the\meqno}
          \else\ifnum\appno>0
            \eqno({\rm\romappno}.\the\meqno)\xdef#1{{\rm\romappno}.\the\meqno}
          \else
            \eqno(\the\meqno)\xdef#1{\the\meqno}
          \fi
        \fi
\global\advance\meqno by1 }


\global\newcount\refno
\global\refno=1 \newwrite\reffile
\newwrite\refmac
\newlinechar=`\^^J
\def\ref#1#2{\the\refno\nref#1{#2}}
\def\nref#1#2{\xdef#1{\the\refno}
\ifnum\refno=1\immediate\openout\reffile=refs.tmp\fi
\immediate\write\reffile{
     \noexpand\item{[\noexpand#1]\ }#2\noexpand\nobreak.}
     \immediate\write\refmac{\def\noexpand#1{\the\refno}}
   \global\advance\refno by1}
\def\semi{;\hfil\noexpand\break ^^J}
\def\nl{\hfil\noexpand\break ^^J}
\def\refn#1#2{\nref#1{#2}}

\def\ann#1#2#3{{\it Ann. Phys.} {\bf {#1}} (19{#2}) #3}
\def\cmp#1#2#3{{\it Commun. Math. Phys.} {\bf {#1}} (19{#2}) #3}

\def\jmp#1#2#3{{\it J. Math. Phys.} {\bf {#1}} (19{#2}) #3}

\def\ijtp#1#2#3{{\it Int.  J.  Theor.  Phys.} {\bf {#1}} (19{#2}) #3}

\def\np#1#2#3{{\it Nucl.  Phys.} {\bf B{#1}} (19{#2}) #3} 
 
\def\plA#1#2#3{{\it Phys.  Lett.} {\bf {#1}A} (19{#2}) #3}

\def\prD#1#2#3{{\it Phys.  Rev.} {\bf D{#1}} (19{#2}) #3}

\def\prp#1#2#3{{\it Phys.  Rep.} {\bf {#1}C} (19{#2}) #3}


{

\refn\MaF
{V.P. Maslov and M.V. Fedoriuk, 
\lq\lq Semiclassical Approximation in Quantum \nl
Mechanics\rq\rq, 
D. Reidel Publ., London, 1981}

\refn\GY
{I.M. Gel'fand and A.M. Yaglom, \jmp{1}{60}{48}}

\refn\LS
{S. Levit and U. Smilansky,
{\sl Proc. Amer. Math. Soc.} {\bf 21} (1977) 299;
\ann{103}{77}{198}}

\refn\S
{L.S. Schulman, in \lq\lq Functional Integration
and its Applications\rq\rq, A.M. Arthurs, ed., 
Clarendon Press, Oxford, 1975}

\refn\Schulman
{L.S. Schulman, \lq\lq Techniques and Applications of
Path Integration\rq\rq, John Wiley \& Sons, New York, 1981}

\refn\DM
{C. DeWitt-Morette, \ann{97}{76}{367}}

\refn\DMMN
{C. DeWitt-Morette, A. Maheshwari and B. Nelson, 
\prp{50}{79}{256}}

\refn\DMNZ
{C. DeWitt-Morette, B. Nelson and T.-R. Zhang,
\prD{28}{83}{2526}}

\refn\DV
{G. Dangelmayr and W. Veit,
\ann{118}{79}{108}}

\refn\Berry
{M.V. Berry, 
{\sl Adv. Phys.} {\bf 25} (1976) 1}

\refn\Witten
{E. Witten, \np{149}{79}{285}}

\refn\Jevicki
{A. Jevicki, \prD{20}{79}{3331}}

\refn\CH
{R. Courant and D. Hilbert, 
\lq\lq Methods of Mathematical Physics\rq\rq, 
Interscience Publishers, New York, 1953}

\refn\MF
{P.M. Morse and H. Feshbach,
\lq\lq Methods of Theoretical Physics\rq\rq, 
McGraw-Hill Book Company, New York, 1953}

\refn\Milnor
{J. Milnor, \lq\lq Morse Theory\rq\rq, 
Princeton University Press, Princeton, 1963}

\refn\Souriau
{J.-M. Souriau, in \lq\lq Group Theoretical Methods
in Physics\rq\rq, A. Janner, T. Janssen and M. Boon, eds., 
Lecture Notes in Physics, {\bf 50}, Springer-Verlag, Berlin, 1976}

\refn\Horv
{P.A. Horv\'{a}thy, \ijtp{13}{79}{245}}

\refn\C
{B.K. Cheng, \plA{101}{84}{464}}

\refn\FH
{R.P. Feynman and A.R. Hibbs, 
\lq\lq Quantum Mechanics and Path Integrals\rq\rq, 
McGraw-Hill, New York, 1965}

\refn\H
{G.A. Hagedorn, \cmp{71}{80}{77}}

\refn\TT
{S. Tanimura and I. Tsutsui, 
\ann{258}{97}{137}}





}

\def\ve{\vfill\eject}




\pageheight{23cm}
\pagewidth{14.8cm}
\hcorrection{0mm}
\magnification= \magstep1
\def\bsk{%
\baselineskip= 16.8pt plus 1pt minus 1pt}
\parskip=5pt plus 1pt minus 1pt
\tolerance 6000


\null


{
\leftskip=100mm
\hfill\break
KEK Preprint 98-161
\hfill\break
hep-th/9810156
\hfill\break
\par}

\smallskip
\vfill
{\baselineskip=18pt

\centerline{\bigbold 
Quantum Caustics for Systems with Quadratic Lagrangians}
 
\vskip 30pt

\centerline{
\smc 
Kenichi Horie,
\quad 
Hitoshi Miyazaki
\quad {\rm and} \quad 
Izumi Tsutsui\note
{E-mail:\quad tsutsui@tanashi.kek.jp}
}

\vskip 5pt

{
\baselineskip=13pt
\centerline{\it 
Institute of Particle and Nuclear Studies}
\centerline{\it 
High Energy Accelerator Research Organization (KEK),
Tanashi Branch}
\centerline{\it Tokyo 188-8501, Japan}
}

\vskip 15pt

\centerline{
\smc Shogo Tanimura\note
{E-mail:\quad tanimura@kuamp.kyoto-u.ac.jp}
}

\vskip 5pt

{
\baselineskip=13pt
\centerline{\it
Department of Applied Mathematics and Physics}
\centerline{\it 
Graduate School of Informatics, Kyoto University}
\centerline{\it Kyoto 606-8501, Japan}
}

\vskip 60pt

\abstract{%
{\bf Abstract.}\quad
We study caustics in classical and 
quantum mechanics for systems with
quadratic Lagrangians of the form
$L = {1 \over 2} \dot x^2 - {1 \over 2} \lambda(t) x^2 - \mu(t) x$.
We derive a closed form of the transition 
amplitude on caustics 
and discuss their physical implications 
in the Gaussian slit (gedanken-)experiment.
Application to the quantum mechanical rotor
casts doubt on the validity of Jevicki's correspondence 
hypothesis which states that in quantum mechanics, 
stationary points (instantons) arise as  
simple poles.
}

\bigskip
{\ninepoint
PACS codes: 02.30.Hq; 03.65.-w; 03.65.Sq \hfill\break 
}


\pageheight{23cm}
\pagewidth{15.7cm}
\hcorrection{-1mm}
\magnification= \magstep1
\def\bsk{%
\baselineskip= 15.2pt plus 1pt minus 1pt}
\parskip=5pt plus 1pt minus 1pt
\tolerance 8000
\bsk


\ve

\secno=1 \meqno=1 


\centerline{\bf 1. Introduction}
\medskip

Semiclassical approximation is a powerful and perhaps 
the most commonly used approach to quantum mechanics for 
exploiting classical mechanics 
based on the idea that the former 
may be realized by supplementing the latter properly
(see, {\it e.g.}, [\MaF]).  
In the path-integral language, it asserts that
the transition amplitude between two arbitrarily 
given points may be approximated by 
summing up fluctuations around the classical path
connecting the two points.  Here, the existence of
the classical path is assumed, not guaranteed, 
in the first place.

Caustics occur when this assumption breaks down.  
More precisely, 
when a family of classical trajectories focuses,
the envelope of the trajectories forms a focal 
region called caustics.  In one dimension, 
the region becomes a focal point, and hence a classical path
connecting generic initial and final 
points $a$, $b \in \R$ exists
if and only if $b$ is the focal point 
specified by $a$.
Given the action $I$ of the system, 
this happens when the second
variation of the action $\delta^2 I$ along a classical
path starting from $a$
vanishes, and as such their analysis
is purely classical and constitutes a branch of 
the Sturm-Liouville problem.  
In particular, 
if we confine ourselves to the 
action $I[x] = \int_0^T dt\, L$ for
a finite time interval $[0, T]$ with
the quadratic Lagrangian\note{Note that 
any Lagrangian at most quadratic in position and
velocity can be brought into this form
by partial integration.  
In this paper we use the dot to denote 
time derivative $\dot x = dx/dt$, and
put both the mass $m$ of the particle
and the Planck constant $\hbar$ (except section 3.2)
unity for convenience.}
$$
  L = {1 \over 2}\dot{x}^2
     -{1 \over 2}\lambda(t)\,x^2
     -\mu(t)\,x\ ,
\eqn\lagrange
$$
then $\delta^2 I =0$ is equivalent to 
the condition that a solution $u$ obeying the homogeneous
equation of motion
$\ddot u(t) + \lambda(t)\, u(t) = 0$ 
with $u(0) = 0$ vanish at $t = T$. 

In quantum mechanics, 
the semiclassical approximation for the quadratic system 
is known to be
exact, and in terms of the Morse index $m(\lambda)$ and 
the action $I[x_{\rm cl}]$ evaluated for a 
classical path $x_{\rm cl}(t)$,
the amplitude for the transition between the
two points $a$, $b$ reads [\GY, \LS]
$$
K(b,T;a,0)
       = \left( {1 \over {2 \pi i \vert u(T) \vert}} 
\right)^{1\over 2}\,
       e^{i I[x_{\rm cl}] - {{i\pi}\over 2} m(\lambda) }.
\eqn\semiexact
$$
Thus the amplitude becomes singular when the harmonic 
potential $\lambda$ admits the solution $u$ to become
$u(T) = 0$, that is, when caustics occur.
The appearance of the singularity suggests that, on caustics,
a generalized prescription is required for  
semiclassical approximation to treat the cases
where the classical path does not exist.
Such a prescription has been devised
in the path-integral framework 
[\S, \Schulman] (cf.~section 3.1 of this paper).
Strictly speaking, the singularity does
not arise in actual physical processes, because
there usually exist higher order terms (cubic, quartic 
{\it etc.}~in $x$) in the Lagrangian, 
or even if the quadratic form (\lagrange) 
provides a reasonably accurate description of the system, 
the potential $\lambda$ may not allow the solution $u$ to vanish
at $t = T$, which is generically the case.        
Nonetheless, physical
phenomena pertaining to caustics do arise, and the quantum 
analysis for such phenomena occurring in the presence of
higher order terms has been carried out 
in the path-integral by 
Schulman [\S] and DeWitt-Morette [\DM] (see also 
[\Schulman, \DMMN, \DMNZ, \DV]; for physical aspects of
caustic phenomena in wave theory and the relation 
to Thom's theorem, see, {\it e.g.,} [\Berry]).

In this paper,
we wish to provide a study of quantum caustics characteristic to
the quadratic Lagrangian (\lagrange) for a generic potential
$\lambda$ and an external force $\mu$.
The aim of the study is three-fold: 
(i) to present a basic but self-contained result on 
classical caustics which can be used for a fuller
analysis of caustics for general Lagrangians
(because caustics are characterized by the quadratic 
part of the Lagrangian),
(ii) to obtain a formula for the 
transition kernel $K(b,T;a,0)$ which covers the case of caustics 
and introduce the notion of a quantum Jacobi field to 
discuss the physical 
implications of caustics at the quantum level,
and (iii) to apply the result to examine
Jevicki's correspondence hypothesis, which we now explain below.

In an attempt to resolve certain discrepancies between
instantons in QCD and the $1/N$ expansion, 
it was argued by Witten [\Witten] in
the two dimensional $\Complex P^N$ nonlinear sigma model
(which is a prototype of QCD) that 
classical instantons are eliminated by
quantum fluctuations and
disappear at the quantum level.  To this assertion
Jevicki [\Jevicki] contended --- under a few but crucial 
assumptions --- that instantons do not disappear
but show up in the form of simple poles 
rather than stationary points at the quantum level. 
His point was illustrated by
the example of the quantum mechanical rotor, {\it i.e.}, 
a free particle on the circle $S^1$, where
the correspondence
$$
\hbox{stationary points}\quad \Longleftrightarrow \quad 
\hbox{simple poles} 
\eqn\hypo
$$
may become transparent.  Embedded in the plane $\R^2$,
the rotor can be transformed into a two dimensional
harmonic oscillator with an arbitrary potential $\lambda$,
where our result for the quadratic Lagrangian
is applicable.  Our detailed analysis will 
show that, even in this toy model,
the assumptions made in [\Jevicki] cannot be justified and, 
therefore, the correspondence hypothesis (\hypo) cannot 
be sustained.  In short, there is no firm reason for 
rejecting Witten's assertion.   

The plan of the paper is as follows.  
In section 2 we provide a full account of
classical caustics associated with 
the quadratic Lagrangian (\lagrange).  To render
the classical results most convenient for later use, 
we present them
in a mathematical style.  Readers who wish to
see only the discussion of quantum caustics may skip
this section.  Section 3 is devoted to quantum caustics,
where we first derive the kernel formula which admits the case
caustics and then discuss the quantum effect associated with
caustics by a simple Gaussian slit (gedanken-)experiment.
Jevicki's correspondence hypothesis is examined 
in section 4.  Section 5 contains 
our conclusion and discussions.
An appendix is provided at the end 
to supplement the argument of section 3.

\ve

\secno=2 \meqno=1 


\centerline{\bf 2. Classical Caustics}
\medskip

In this section we review and summarize various
aspects of the classical motion of a point
particle governed by a quadratic Lagrangian.
Our discussions focus especially on the
occurrence of caustics and its consequences in the
framework of classical physics.
Part of the material presented is well-known, see
{\it e.g.}, [\Schulman], but is included here
together with numerous other facets of the issue.
In this way the following discussions illustrate
the phenomenon of caustics in classical physics as
opposed to that in quantum realm in later sections
and further furnish a mathematical basis of
the discussions in the following sections on quantum
mechanics on $S^1$.  First we consider the 
Jacobi equation, which in the
case of quadratic Lagrangian is the homogeneous part
of the equation of motion, and discuss some general
properties of its solutions.
Next we concentrate on the phenomenon of caustics and
discuss two of its main characteristics, the one being
the constant stretching
factor between the final and the initial points, and
the other being the Morse index.

\medskip
\noindent
{\bf 2.1. Quadratic Lagrangian and Jacobi equation}

Let us consider a point particle governed by the
quadratic Lagrangian (\lagrange), whose 
equation of motion reads
$$
 A_{\lambda}\, x(t) = \mu(t)\ , \qquad  \hbox{where}
\quad 
  A_{\lambda} :=%
  - \left[ {{d^2}\over{dt^2}} + \lambda(t) \right] \ .
\eqn\motionfull
$$
{}For the discussion of caustics it is useful to
consider the Jacobi field $ J(p,t) $
(see {\it e.g.}, [\Schulman])
which describes the spread of paths from an initial
point when varying the momentum.  Given a one-parameter
family of classical paths $x(p, t)$ 
characterized by the initial momenta 
$\dot x(p, t) = p$ at $t = t_0$, 
it is defined by
$$
  J(p,t) := {\partial x(p,t) \over \partial p}\ .
\eqn\jacobi
$$
It satisfies $J(p,t_0)=0$ by definition and solves 
the Jacobi equation, which in case of the
quadratic Lagrangian (\lagrange) is the
homogeneous part of (\motionfull). 
Hence, the Jacobi equation can itself be interpreted
as the force-free equation of motion,
$$
  A_{\lambda}\, u(t) = 0\ ,
\eqn\motion
$$
and as such contains a wider class of solutions
(not necessarily zero at $t=0$) which may be added to a
solution curve of the full equation of motion
(\motionfull) to yield other extremum paths.
The investigation of the solution curves to (\motion)
not only leads to a characterization
of caustics via the Jacobi field itself, but reveals
further properties of the caustics, see in this context also
[\DM].
In particular, the relationships between two linearly
independent solutions of the homogeneous differential
equation of second order (\motion) help us to clarify the
notion of the stretching factor and the index 
which will be introduced later on.

As is well known, given initial data $x(t_0)=x_0$ and
$\dot{x}(t_0)=\dot{x}_0$ at $t_0$, the differential
equation (\motion) possesses a unique solution.
If $\lambda(t)$ is continuous, the solution is of
class ${\cal C}^2$, {\it i.e.}, has continuous
derivatives of order two, and when $\lambda(t)$ is only
piecewise continuous, the solution curve is at least
${\cal C}^1$.
In what follows, we shall use frequently the following

\medskip
\noindent
{\bf Lemma 1.}
{\it Let $v(t)$ and $u(t)$ be solutions of the
differential equation (\motion), and let $t_1$ and $t_2$
be two times. Then we have}
$$
   v\dot{u} \Big\vert^{t_2}_{t_1}
 = \dot{v}u \Big\vert^{t_2}_{t_1} \ .
\eqn\idint
$$

\medskip
\noindent
{}For the proof, we observe from (\motion) that
$$
  v\dot{u} \Big\vert^{t_2}_{t_1}
   = \int_{t_1}^{t_2}dt {{d}\over{dt}}(v\dot{u})
   = \int_{t_1}^{t_2}dt {{d}\over{dt}}(\dot{v}u)
   + \int_{t_1}^{t_2}dt (- \ddot{v}u + v\ddot{u} ) 
   = \dot{v}u \Big\vert^{t_2}_{t_1} \, .
\eqn\preqone
$$

There are two, linearly independent
solutions to (\motion), which form the space
of solutions associated with (\motion).
In the case of constant positive 
$\lambda := \omega^2$,
a basis for the 2-dimensional solution space of (\motion)
may be given by $\{\sin(\omega t), \cos(\omega t)\}$.
We shall remark here that, even for generic $\lambda(t)$,
any two linearly independent solution curves behave like
sine and cosine if they are to possess zeros.
To this end, first we state the following lemma,
which is easily verified by direct differentiation:

\medskip

\noindent
{\bf Lemma 2.}
{\it Let $u(t)$ and $v(t)$ be two linearly independent
solutions of (\motion) and let $u(t_0)=0$. Then
$v(t_0) \ne 0$, and $u$ may be expressed as}
$$
  u(t) = \dot{u}(t_0) \cdot v(t)\, v(t_0)
         \int_{t_0}^t{{ds}\over{(v(s))^2}}
\eqn\udef
$$
{\it as long as $v$ is non-zero.}

\medskip
\noindent
{}Formula (\udef) can be used to create a linearly
independent solution with prescribed zero position and
initial velocity to a given solution curve.
Before embarking on further relationships between $v$
and $u$, we show

\medskip
\noindent
{\bf Lemma 3.}
{\it Let $v(t)$ be a (non-trivial) solution of (\motion)
and let $\tau_1$ be a  time such that $v(\tau_1)=0$. Then
either $\dot{v}(\tau_1) \ne 0$ or else $\lambda$ has a
pole at $\tau_1$. In particular, if $\lambda$ is piecewise
continuous, then the zeros of $v$ are separated. }

\medskip
\noindent
{\it Proof.} Let $t_0$ be a time where $v(t_0) \ne 0$, and
define $u(t)$ according to {Lemma 2} with $\dot{u}(t_0) = 1$.
With the help of {Lemma 1} applied to time $t_0$ and
$\tau_1 - \epsilon$ with $\epsilon > 0$, we have
$$
     v(\tau_1-\epsilon)\dot{u}(\tau_1-\epsilon) - v(t_0)
  = \dot{v}(\tau_1-\epsilon)u(\tau_1-\epsilon) \, .
\eqn\nor
$$
Thus, by letting $\epsilon$ go to zero, $\dot{v}(\tau_1)=0$
is possible only if $u(\tau_1) = \infty$ or
$\dot{u}(\tau_1) = \infty$, which both imply that $u$ can
not be ${\cal C}^1$ at $\tau_1$. Therefore, $\lambda$ is neither
continuous nor even piecewise continuous at $\tau_1$,
proving the assertion. {\it Q.E.D.}

\medskip
Let us henceforth assume that $\lambda$ be piecewise 
continuous or, if not, be at least such
that whenever $v(\tau_1) = 0$, then $\dot{v}(\tau_1) \ne 0$.
We then have the following

\medskip
\noindent
{\bf Lemma 4.}
{\it In the situation of {Lemma 2} let $\tau_1$ be the first
zero position of $v$ beginning with $t_0$. Then the formula
(\udef) for $u$ is well-defined at $\tau_1$ and its value is
given by $u(\tau_1)= - \dot{u}(t_0) v(t_0)/ \dot{v}(\tau_1)$. }

\medskip
\noindent
{\it Proof.} Since $u$ is a solution of (\motion) and
$\lambda$ is piecewise continuous, any possible singularity
of $u$ stemming from its definition (\udef) can be restored,
and the value of $u$ at $\tau_1$ can be calculated with the
help of {Lemma 1}.
One can check this also directly from (\udef) in the following
way. Let $\epsilon_0$ and $\epsilon$ be positive numbers,
such that $0 < \epsilon < \epsilon_0 \ll 1$.
Now if $v(\tau_1)=0$, then from (\motion) it follows
$\ddot{v}(\tau_1)=0$. Thus in the interval
$[\tau_1-\epsilon_0,\tau_1]$ $v$ can be approximated as
$v(t)=\dot{v}(\tau_1)(t-\tau_1) + {\cal O}((t-\tau_1)^3)$.
Keeping $\epsilon_0$ fixed, but letting $\epsilon$ go to
zero we obtain
$$
\eqalign{
  u(\tau_1-\epsilon)
  &= \dot{u}(t_0) \left(-\epsilon\dot{v}(\tau_1)
           +{\cal O}(\epsilon^3)   \right)\, v(t_0)
     \left[ \int_{t_0}^{\tau_1-\epsilon_0}{{ds}\over{(v(s))^2}}
           +\int_{\tau_1-\epsilon_0}^{\tau_1-\epsilon}
             {{ds}\over{(v(s))^2}}
     \right] \cr
  &= {\cal O}(\epsilon)
    + \dot{u}(t_0) \left(-\epsilon\dot{v}(\tau_1)
           +{\cal O}(\epsilon^3)   \right) v(t_0)
           {1 \over{(\dot{v}(\tau_1))^2}} 
              \int_{\epsilon}^{\epsilon_0}dr
                \left\{ {{1}\over{r^2}}
                      \left( 1+{\cal O}(r^2) \right)
                \right\} \cr
  &= {\cal O}(\epsilon)
    -{{\dot{u}(t_0) v(t_0)}\over{\dot{v}(\tau_1)}} \, ,
}
$$
proving our assertion. {\it Q.E.D.}

\medskip

In the following proposition we now make precise in what
sense two linearly independent solutions $v$ and $u$
behave like sine and cosine for the case when $\lambda$
is a positive constant number.  For the 
solution curves of the differential equation
(\motion) to behave actually like sine or cosine, they must
have zeros. This is certainly not the case for example for
the constant $\lambda \equiv -c^2 < 0$ with one possible
solution curve being $\exp(ct)$. On the other hand, as we
shall see further below, for the case of critical $\lambda$, 
the solution curves necessarily have zeros.
Thus for the following proposition we take it for granted
that the solution curves indeed possess zeros, which are
separated according to {Lemma 3}.

\medskip
\noindent
{\bf Proposition 5.}
{\it Let $u$ and $v$ be two linearly independent solutions
of (\motion). Let ..., $t_{-1}$, $t_0$, $t_1$, $t_2$, ... be
the zero positions of $u$, and ..., $\tau_{-1}$, $\tau_0$,
$\tau_1$, $\tau_2$, ... be those of $v$.
Assume $t_0 \in (\tau_0,\tau_1)$, $v(t_0) > 0$ and
$\dot{u}(t_0) > 0$ for definiteness (other cases of signs
may be treated analogously).
Then $u$ is positive (negative) on $(t_i,t_{i+1})$ for $i$
even (odd), $v$ is positive (negative) on
$(\tau_i,\tau_{i+1})$ for $i$ even (odd), and the zero
positions of $v$ and of $u$ alternate such that }
$$
  ...< \tau_{-1} < t_{-1} < \tau_0 < t_0 < \tau_1 < t_1 <...
\eqn\tseq
$$
\topinsert
\vskip 1.5cm
\let\picnaturalsize=N
\def\picsize{6cm}
\def\picfilename{fig1.epsf}
\ifx\nopictures Y\else{\ifx\epsfloaded Y\else\input epsf \fi
\global\let\epsfloaded=Y
\ifx\picnaturalsize N\epsfxsize \picsize\fi
\hskip 2.5cm\epsfbox{\picfilename}}\fi
\vskip 0.5cm
\abstract{%
{\bf Figure 1.}
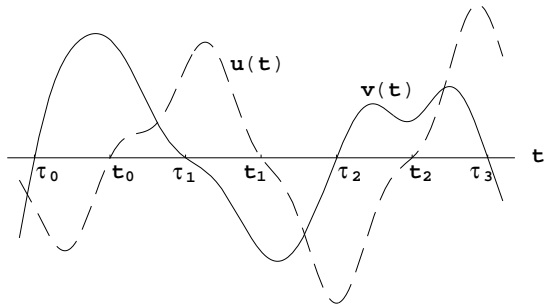
A schematic picture of how two linearly independent
solutions, $v$ and $u$, behave under a generic potential $\lambda$.
}
\endinsert
\medskip
\noindent
{\it Proof.} First of all, the sign of $u$ (and likewise of
$v$) must alternate as one passes the zeros, since at each
zero position the slope of $u$ is non-zero by {Lemma 3}.
We show first that the zero position $t_1$ following $t_0$ lies
in $(\tau_1,\tau_2)$. Since $v$ is positive on $(\tau_0,\tau_1)$,
$u$ may be defined by (\udef) and is therefore strictly positive
on $(t_0,\tau_1)$. Moreover, since the derivative of $v$ at the
zero position $\tau_1$ is non-zero ({Lemma 3}), and since $v$
becomes negative when it crosses $\tau_1$, we must have
$\dot{v}(\tau_1) < 0$. Then, according to {Lemma 4}, $u$ is
strictly positive also at $\tau_1$, which implies $\tau_1 < t_1$.
By applying {Lemma 1} to the interval $[\tau_1,\tau_2]$ one has
$u(\tau_2)\dot{v}(\tau_2)=u(\tau_1)\dot{v}(\tau_1)
=-\dot{u}(t_0) v(t_0)<0$,
the second equality following from {Lemma 4}. Since, by
definition, $v$ has a zero at $\tau_2$ and becomes positive
thereafter (until the next zero at $\tau_3$),
its derivative $\dot{v}(\tau_2)$ is positive, which implies
that $u(\tau_2) < 0$, {\it i.e.}, $u$ must have a zero
at some $t_1 < \tau_2$.
Having proven $\tau_1 < t_1 < \tau_2$, we may proceed by
re-defining $u$ on the interval $[\tau_1,\tau_2]$ analogously
to (\udef),
$$
  u(t) =
  \dot{u}(t_1)\cdot v(t)\, v(t_1)
  \int_{t_1}^{t}{{ds}\over{(v(s))^2}}\ .
\eqn\udefb
$$
The above reasoning for $t_1$ may be applied again to yield
$\tau_2 < t_2 < \tau_3$ (see Fig.1). Proceeding similarly for
other $t_i$, one may prove (\tseq), {\it Q.E.D.}

\medskip

We remark that the formula (\udef) valid on $[\tau_0,\tau_1]$
can be extended to the interval $[\tau_i,\tau_{i+1}]$ by
simply replacing the lower integration boundary $t_0$ in (\udef)
to $t_i$. To see this, one uses (\udefb) and
$\dot{u}(t_1)v(t_1) = \dot{u}(t_0) v(t_0)$ (by Lemma 1).

\medskip
\noindent
{\bf 2.2. Critical potential and its characteristics}

We now consider the full equation of motion (\motionfull)
on a fixed time interval $[0,T]$, and examine the phenomenon
of caustics.  First we recall that, if $u$, $v$ are 
the two linearly independent solutions of the homogeneous part
(\motion) obeying the initial conditions, 
$$
u(0) = 0, \qquad \dot u(0) = 1, \qquad 
v(0) = 1, \qquad \dot v(0) = 0,
\eqn\inbond
$$
then the general solution of the full equation (\motionfull)
is given by
$$
x(t) = \alpha\, v(t) + \beta\, u(t) + s(t) \ ,
\eqn\gensolution
$$
where $s$ is a special solution of (\motionfull) which
we we take to be one satisfying $s(0) = 0$.  (We could
further specify by requiring, say, 
$\dot s(0) = 0$ but this is not important
at the moment.)  The constants $\alpha$, $\beta$ are determined
>from the initial position $x(0)$ and velocity
$\dot x(0)$, respectively.  Note that the solution
$u$ may be obtained from
the Jacobi field (\jacobi) by $u(t) = J(p, t)/\dot J(p, 0)$.

Caustics occur when the solution
$u(t)$, or the Jacobi field $J(p, t)$, 
vanishes at $t = T$.
If this happens then it follows that the final
position $x(T) = \alpha\, v(T) + s(T)$ does not depend
on the initial velocity $\dot x(0)$ (which is also
obvious from the definition of the Jacobi field).    
More precisely, we have

\medskip
\noindent
{\bf Lemma 6.}
{\it Given a time interval $[0,T]$, 
suppose that the Jacobi field $J(p,t)$ beginning at $t=0$ 
vanishes at $T$.  Then there exists a solution
of the full equation of motion
(\motionfull) satisfying the Dirichlet boundary condition,
$$
  x(0) = a, \qquad x(T) = b
\eqn\vab
$$
if and only if
$$
  b = v(T)\cdot a + s(T)\ ,
\eqn\klambda
$$
where $v(T)$ and $s(T)$ are the final values of the 
aforementioned solutions.  Further, 
for the final point $b$ given in (\klambda)
there are infinitely many
solutions satisfying (\vab), and each solution is uniquely
characterized by the initial velocity 
$\dot{x}(0)$, which can take any value. }

\medskip
\noindent
{\it Proof.}
In the general solution (\gensolution) 
the Dirichlet boundary condition (\vab) implies that
$\alpha = a$ while $\beta$ remains arbitrary since
we now have $u(T) = 0$.
Thus whatever the initial velocity (which could determine
$\beta$) may be, the final point $x(T)$ turns out to 
be given by (\klambda).
Since $\beta$ is left arbitrary, there are infinitely many
solutions satisfying (\vab) with different initial
velocities, $\dot x(0) = \beta + \dot s(0)$.  {\it Q.E.D.}

\medskip
Thus, if $J(p, T) = 0$,
the different
solution curves which spread out of the initial point $a$
are all focused in one final point $b$.
This phenomenon can be considered as a special case of
caustics in geometric optics, see {\it e.g.}, [\Schulman].
In this paper we fix the interval $[0, T]$ and study the
implications of caustics that arise under a generic
potential $\lambda$, and to this end we shall 
make the following

\medskip
\noindent
{\bf Definition 7.} If the Jacobi field $J(p,t)$ vanishes
at $t=T$, 
then the potential $\lambda$ shall be called {\it critical},
otherwise {\it non-critical}.

In case of critical potentials, 
it is useful to introduce the constant
$$
k(\lambda) := {{v(T)}\over{v(0)}}\ ,
\eqn\stretch
$$
>from a solution $v$ 
satisfying (\motion) with $v(0) \ne 0$,
which gives the stretching factor during the period $[0, T]$.
Clearly, this constant is independent of the choice of 
the solution $v$ and hence determined solely by 
the critical potential $\lambda$.
Given an initial point $x(0)=a$, 
the final point (\klambda) to which all
solution curves are focused on now reads 
$x(T) = k(\lambda)\,a + s(T)$, and 
is called the {\it conjugate} point to $a$ 
or {\it focal} point.
The simplest example of caustics arises in the harmonic
oscillator, where $\lambda = \omega^2$ is a positive
constant at one of the frequencies
$\omega = {{n\pi}\over{T}}$ for $n = 1$, 2, $\ldots$.
The stretching factor associated with the potential
is then $k(\lambda) = (-1)^n$, and it is easy
to see that there are infinitely many
solutions $x(t)$ arriving at 
the conjugate point $(-1)^n a$.

The following Lemma shows how much the situation of 
a critical potential differs from that of non-critical one.

\medskip
\noindent
{\bf Lemma 8.} {\it If $\lambda$ is non-critical,
there exists a unique solution of (\motionfull) for the
Dirichlet boundary condition (\vab) with 
any $a$ and $b$. }

\medskip
\noindent
{\it Proof.}
Since $\lambda$ is non-critical we have 
$u(T) \ne 0$.  Then, 
for any $a$, $b \in \R$, the Dirichlet boundary condition
(\vab) determines uniquely the constants $\alpha$, $\beta$ 
in the general solution (\gensolution)
to yield
$$
x(t) = a\, v(t) + 
{{ b - a\, v(T) - s(T)}\over {u(T)}}\, u(t) + s(t) \ .
\eqn\helpa
$$
{\it Q.E.D.}

\medskip
We note in passing that
a generic potential is non-critical, and this can be seen
roughly as follows.
Consider a non-critical potential and a solution curve of
the force-free equation of motion (\motion) with
initial value $0$ and initial velocity $1$, which thus
does not vanish at the final time $T$.
If the potential is varied in an infinitesimal manner,
then this solution curve also varies very little,
implying that the final value remains nonzero for such
small variations. This implies that the set of non-critical
potentials is open in an appropriate topology. Further, by
the same reasoning, if the potential is critical, then in
general a small variation suffices to make the potential
non-critical. Thus, again in an appropriate topology, which
we do not specify here but is easy to find, the set of
critical potentials is nowhere dense.

The equation of motion (\motionfull) can be
derived from the action functional
$ I[x] = \int_0^T dt \, L $ with the Lagrangian $L$ in
(\lagrange), 
which now acts on the
space of paths $x(t)$ with some given boundary condition
$x(0)=a$, $x(T)=b$.
If the potential is non-critical, then there exists a
unique classical path from $a$ to $b$, denoted here by
$x(t)$, at which the action becomes a minimum.
Suppose we vary the potential $\lambda$ and with it
the classical path $x(t)$ and its action
$I[x]$ in a small neighbourhood of a critical
potential $\bar\lambda$ with
$a$ and $b$ fixed, wherein we only consider
non-critical potentials.
If $\lambda$ approaches $\bar\lambda$, two cases may occur:
either $b$ happens to be 
conjugate to $a$ under $\bar\lambda$, {\it i.e.},
$b = a\, k(\bar\lambda) + s(T)$, in
which case there exists a classical path also for the critical
potential, and the action of course remains finite,
or else $b$ is {\it not} conjugate point to $a$, 
in which case
there exists no classical path at the caustic, and the
action becomes infinite, as the following lemma shows.

\medskip
\noindent
{\bf Lemma 9.} {\it In the situation of a non-critical
potential $\lambda$ approaching a critical potential
$\bar\lambda$, the
minimum action $I[x]$ of the classical path
$x(t)$ between two given fixed boundary points
tends to infinity if the final point is not a conjugate
one of the initial point for the critical potential
$\bar\lambda$. }

\medskip
\noindent
{\it Proof.}
Consider the classical path satisfying $x(0)=a$ and $x(T)=b$
given in the form (\helpa), where we specify 
the initial velocity of the special
solution $s$ as 
$\dot{s}(0)=0$ for definiteness. 
Denoting $\tilde x(t):= a\, v(t) + s(t)$, and
using the differential equations satisfied by $s$ and $u$,
we obtain the identity,
$$
\int_0^T dt\, \mu(t)\, x(t) = 
\int_0^T dt\, \mu(t)\, \tilde x(t)
- {{b - \tilde x(T)}\over{u(T)}} 
\bigl( \dot s(T) u(T) - s(T) \dot u(T) \bigr)\ .
\eqn\lemident
$$
With the help of this identity and Lemma 1 applied to
$u$ and $v$ of (\helpa), 
we find that the action becomes
$$
  I[x]
  =-{1 \over 2}\int_0^T dt\, \mu(t)\, \tilde x(t)
   +{1 \over 2}\left( 2b-\tilde x(T) \right)
    \dot{\tilde x}(T)
   +{1 \over 2}\left( b -\tilde x(T) \right)^2
    {\dot{u}(T) \over  u(T)} \, .
\eqn\helpb
$$
As $\lambda$ approaches $\bar\lambda$, we observe that
$u(T) \rightarrow 0$,
$\dot{u}(T) \rightarrow 1/k(\bar\lambda)$,
$\tilde x(T) \rightarrow k(\bar\lambda)\,a + s(T)$, while
$\dot{\tilde x}(T)$ remains finite since 
its initial velocity $\dot{\tilde x}(0) = 0$
is fixed for all $\lambda$. 
The assertion then follows
>from the assumption, $b \ne k(\bar\lambda)\,a + s(T)$.  {\it Q.E.D.}

\medskip
\noindent
An important point to note is
that the sign of the divergence $I[x] \rightarrow \pm \infty$
depends on the sign of the product $u(T)\,k(\lambda)$
in the limit $\lambda \rightarrow \bar\lambda$, and this depends
on how $\lambda$ approaches $\bar\lambda$, not just on $\bar\lambda$
it is approaching.

If $\lambda$ is critical, then besides the
stretching factor $k(\lambda)$, a further characteristics
is the index of the differential operator $A_\lambda$
viewed as a symmetric bilinear functional, which we now
illustrate (see {\it e.g.},\ section 12 in [\Schulman]).
Consider a generic potential $\lambda$ with external force
$\mu$ and the action functional $I[x]$ for paths $x$ 
with fixed endpoints $x(0)=a$, $x(T)=b$, 
and let $x_{\rm cl}(t)$ be a
classical path (which in case of critical $\lambda$ is
assumed to exist) which obeys the boundary condition. 
To $x_{\rm cl}(t)$ we add a
perturbation $\eta(t)$ with $\eta(0) = \eta(T) = 0$
and expand the action $I[x_{\rm cl}+\eta]$ in terms of
$\eta$, the calculation of which is straightforward,
$$
  I[x_{\rm cl}+\eta]
  = I[x_{\rm cl}]
   +\int_0^T dt\, \eta\, (A_\lambda x_{\rm cl}-\mu)
   +{{1}\over{2}}
     \int_0^T dt\, \eta A_\lambda \eta \ .
\eqn\Iexpand
$$
Since $x_{\rm cl}$ is a classical solution, the second
term linear in $\eta$ on the right hand side vanishes.
The last term on the right describes the
action of a bilinear functional on the space of paths
defined on $[0,T]$ with vanishing boundary values.
The {\it index} of this (non-degenerate)
bilinear functional is then defined by the dimension of the
space on which it is negative definite.
It characterizes the
type of the saddle point of the action functional $I$ 
at the classical path $x_{\rm cl}$.

{}From the consideration of Sturm-Liouville problem the
operator $A_\lambda$ is known to possess a complete set
of orthonormal eigenfunctions $u_n(t)$ [\CH, \MF]:
$$
  A_{\lambda} u_{n}(t) =
  -\left[ {{d^2}\over{dt^2}} + \lambda(t) \right]
  u_{n}(t) = E_n\, u_{n}(t)\ ,
\eqn\eigen
$$
with
$$
  u_n(0) = u_n(T) = 0\, ; \qquad
  \int_0^T dt\, u_n(t)\, u_m(t) = \delta_{nm} .
\eqn\usn
$$
The index is then given by
the number of negative eigenvalues of the
eigenvalue problem (\eigen) for non-degenerate $A_\lambda$
for which no zero mode exists.
It is known that for large $n$
the eigenfunctions $u_n$ approach trigonometric functions
with the corresponding asymptotic eigenvalues
$\big({{n\pi}\over{T}}\big)^2$ [\CH, \MF], which
are thus unbounded from above.
In the following, we assume the eigenvalues to be
bounded from below, which is for example the case
when $\lambda$ is bounded from above.

The bilinear functional $A_\lambda$ becomes degenerated
if there arises a zero mode solution
of (\eigen), {\it i.e.}, a $u_m$ with $E_m = 0$ for some $m$.
The index of $A_\lambda$ may be extended
even to this degenerate case
by saying that it is given by the number of modes with $E_n \le 0$.
By definition, the degeneracy
occurs when the potential $\lambda$ is critical
under the given interval $[0, T]$.
To illustrate this fact from another viewpoint (see {\it e.g.},
pp79 in [\Schulman]) let us consider the initial value problem
$x(0)=x_0$, $\dot{x}(0)=\dot{x}_0$, where we no longer fix
the final time $T$, but let it be variable. Looking for each
$T$ the classical path, the corresponding expansion (\Iexpand),
and the eigenvalue problem (\eigen), with gradually growing $T$,
we observe that the index of the operator $A_\lambda$
changes only at focal points when an eigenvalue goes through
zero, {\it i.e.}, at those values of $T$,
where $\lambda$ restricted on the interval $[0,T]$
becomes critical. The precise statement of this is the following

\medskip
\noindent
{\bf Proposition 10.} {\it Let $u$ be a solution of (\motion)
with initial condition $u(0)=0$. Then the index of the
operator $A_\lambda$ is given by the number of zeros
(focal points) of $u$ on the half-open interval $(0,T]$,
which is called the Morse index.
If $\lambda$ is critical and $v$ a solution of (\motion)
with the initial condition $v(0) \ne 0$, then $v$ must
have a zero on $(0,T)$. The number of zeros of $v$ is
given by the Morse index. }

\medskip
\noindent
{\it Proof.} The first assertion on the index is a special
case of the Morse theorem, cf.\ [\Milnor]. The second
assertion follows directly from {Proposition 5}, since 
between two adjacent focal points of $u$ 
there must lie a zero of $v$. {\it Q.E.D.}

\medskip
Note that, for $\lambda$ critical, the Morse index,
which is characterized by $\lambda$ and is denoted by
$m(\lambda)$, gives the
number of negative eigenvalues of (\eigen) plus one.
We now have the following consequence of the foregoing
discussions on the stretching factor of a critical potential. Let
$\lambda$ be critical, and $u$ be a zero mode
solution of (\motion) with zeros at
$t_0=0$, $t_1$, ..., $t_m=T$, where $m = m(\lambda)$
is the Morse index.
Since a zero mode solution
is unique up to a multiplicative factor, these zeros are
fixed for a given potential $\lambda$.
Let $\lambda_i := \lambda\vert_{[t_{i-1},t_i]}$ be the
restriction of $\lambda$ on the $i$-th interval
$[t_{i-1},t_i]$, $i=1,\, ...,\, m$.
Then the restriction of the zero mode $u$ on each such
interval is trivially a zero mode solution corresponding
to $\lambda_i$. Thus $\lambda_i$ is critical with Morse
index $m(\lambda_i) = 1$. Then one has

\medskip
\noindent
{\bf Lemma 11.} {\it The stretching factor of a critical potential
is the product of the stretching factors of the potentials
restricted to each closed interval between two adjacent
zeros of a zero mode solution, i.e.,}
$$
  k(\lambda)
  = k(\lambda_1) \cdots k(\lambda_m) \, .
\eqn\kkk
$$
{\it Each $k(\lambda_i)$ is negative, and thus the stretching factor
is negative (positive) if the Morse index $m(\lambda)$
is odd (even). }

\medskip
\noindent
{\it Proof.} Let $v$ be a solution of (\motion) linearly
independent from $u$. Then from the proof of {Lemma 6} we
have $k(\lambda_i)=v(t_i)/v(t_{i-1})$ and
$$
k(\lambda) = {{v(T)}\over{v(0)}}
             = {{v(t_m)}\over{v(t_{m-1})}} \cdots
               {{v(t_1)}\over{v(t_0)}}\ .
\eqn\noq
$$
Each stretching factor is negative, since by the proof of
{Proposition 5} between two adjacent zeros of the zero
mode solution $u$ there must lie a zero of $v$, where its
sign changes.  {\it Q.E.D.}

\ve

\secno=3 \meqno=1 


\centerline{\bf 3. Quantum Caustics}
\medskip

In the previous section 
we studied various dynamical aspects of
a particle moving under the influence of 
time dependent harmonic
potentials $\lambda(t)$ and external 
driving forces $\mu(t)$.  
When $\lambda(t)$
is critical, the dynamics of the system
exhibits a number of singular
characteristics at the classical level, notably
in that transitions are allowed
only between conjugate points specified by the stretching factor 
$k(\lambda)$ of the potential.
The aim of this section is to study the quantum dynamics
of the system by looking at the transition amplitude closely.
In particular, for critical $\lambda(t)$,
we shall confirm the known fact [\Schulman, \DM]
that classically forbidden transitions 
continue to be forbidden even at the quantum level.  
The novelty of our result is the closed form
of the path-integral kernel for the 
transition amplitude expressed in terms
of the stretching factor $k(\lambda)$ and the Morse index $m(\lambda)$.
We further discuss the quantum effect in the momentum susceptibility
of a Gaussian wave packet for        
non-critical $\lambda(t)$.

\medskip
\noindent
{\bf 3.1. Transition amplitude and caustics} 

We begin by considering 
the transition amplitude for the system defined
by the quadratic Lagrangian (\lagrange).
Let $\widehat H(t)$ be the Hamiltonian operator
corresponding to the Lagrangian, and $\widehat U(T,0)$
the unitary operator for the time evolution obeying
the Schr\"odinger equation, 
$
(i\partial/\partial t) \widehat U(t,0) 
= \widehat H(t)\, \widehat U(t,0).
$ 
Then, in the path-integral formalism, the
amplitude for the transition from $x = a$ at
$t = 0$ to $x = b$ at $t = T$, 
where $a$ and $b$ are arbitrary
two points on the line $\Real$, is given by
$$
        K (b,T;a,0)
        = \langle b |\, \widehat U(T,0)\, | a \rangle
        = \int_{x(0)=a}^{x(T)=b} {\cal D} x \, e^{i I[x]}.
\eqn\path
$$
In carrying out the path-integration, we need to
take into account the fact that, when $\lambda(t)$
is critical, there may not exist a classical solution
that respects the given boundary condition.
The general case, including the critical one,
may be handled by the following procedure [\Schulman, \S].

First, let $c$ be the endpoint at $t = T$ of a 
classical solution $\bar x_{\rm cl}$ of the equation
of motion (\motionfull) starting from $a$ at $t = 0$, 
{\it i.e.},
$\bar x_{\rm cl}(0) = a$ and $\bar x_{\rm cl}(T) = c$.
Here we put the bar to the solution
in order to emphasize the fact that 
the endpoint value $c$ may not be equal to $b$, 
reserving the notation 
$x_{\rm cl}$ without a bar for 
the actual solution, if any, having
$x_{\rm cl}(T) = b$.  
If $\lambda$ is non-critical, Lemma 8 states that 
$c$ can be chosen arbitrarily and hence the 
solution $x_{\rm cl}(t)$ does exist,
whereas if $\lambda(t)$ is critical, Lemma 6 states that
the endpoint is determined uniquely 
$c = k(\lambda)\,a + s(T)$ 
by the stretching factor of the potential and the endpoint of
the solution $s(t)$.
Now, let us choose a fixed, smooth
function $\rho(t)$ satisfying
$$
        \rho(0) = 0, \qquad \rho(T) = b - c.
\eqn\defrho
$$
Then we may decompose any path 
$x(t)$ connecting the endpoints, $x(0) = a$ and $x(T) = b$, as
$$
        x(t) = \bar x_{\rm cl}(t) + \rho(t) + \eta(t).
\eqn\xdecomp
$$
The function $\eta(t)$ in (\xdecomp), which represents 
the fluctuations, vanishes at both of the ends
$ \eta(0) = \eta(T) = 0 $ and may be expanded  
$ \eta(t) = \sum_n a_n u_n(t) $
in terms of the orthonormal eigenfunctions
in (\eigen) and (\usn).

Using the decomposition (\xdecomp) we find
$$
\eqalign{
        I[ \bar x_{\rm cl} + \rho + \eta ]
        &= 
        I[ \bar x_{\rm cl} ] + I[ \rho ] + I[ \eta ]
        \cr
        & \qquad + 
        \int_0^T dt
        \bigl\{
                \dot{\bar x}_{\rm cl}\, \dot{\rho} +
                \dot{\rho}\, \dot{\eta} +
                \dot{\eta}\, \dot{\bar x}_{\rm cl} 
                - \lambda
                (
                \bar x_{\rm cl}\, \rho +
                \rho\, \eta +
                \eta\, \bar x_{\rm cl} 
                )
        \bigr\}
        \cr
        &= 
        I[ \bar x_{\rm cl} ] + I[ \rho ]\vert_{\mu = 0} 
        + \frac{1}{2} \sum_n E_n a_n^2
        \cr
        &\qquad +
        \dot{\bar x}_{\rm cl}(T) \rho(T)
        + \rho(T) \sum_n a_n \dot{u}_n (T)
        + \sum_n E_n a_n \int_0^T dt \, \rho(t) u_n(t).
}
\eqn\calculus
$$
After the change of the integral variables
${\cal D}x = {\cal D}\eta =  {\cal N}\prod_n da_n$
where $ {\cal N} $ is a Jacobian factor,
the path-integral (\path) becomes a product of Gaussian
integrals over the modes $a_n$.  If $\lambda$ is
non-critical, there exists the 
solution $x_{\rm cl}$ 
and hence we are allowed to choose
$\rho(t) = 0$ identically in (\calculus). 
The Gaussian integrations then gives the standard result,
$$
        K(b,T;a,0) =
        {\cal N}
        \left[ \prod_n E_n \right]^{ -\frac12 }
        e^{ i I[ x_{\rm cl}] }\ ,
\eqn\kernelnonc
$$
which can be shown [\GY, \LS] to be equivalent to (\semiexact).

If, on the other hand, $ \lambda(t) $ is critical, 
then by definition we have $ E_m = 0 $ for some $ m $.   
It follows that the integration over $a_m$, which is now
non-Gaussian, yields the delta-function,
$\delta( \rho(T)\, \dot{u}_m (T) )$.  Hence, for
critical potentials the kernel
formula (\kernelnonc) should be modified into
$$
        K(b,T;a,0) = 
        \sqrt{\frac{2 \pi}{i}} {\cal N} 
        \left[ \prod_{n \ne m} E_n\right]^{ -\frac12 }
        \delta( \rho(T) \,\dot{u}_m (T) )\, e^{iI[x_{\rm cl}]},
\eqn\ckera
$$
where we have set $\rho(t) = 0$ in the phase part of
the kernel (\ckera), which is allowed in the presence of
the delta-function and due to 
the fact $\dot{u}_m (T) \ne 0$.  Since we have now
$$
\rho(T) = b - k(\lambda)a - s(T)\ ,
\eqn\valrho
$$
the kernel (\ckera) already shows that
classically forbidden transitions to non-conjugate points,
$b \ne k(\lambda)a + s(T) $ from $a$, are  
forbidden even quantum mechanically.
We remark at this point that, 
strictly speaking, 
the action $I[x_{\rm cl}]$ in (\ckera) should be
written as $I[\bar x_{\rm cl}]$ since there is no
solution $x_{\rm cl}(t)$ if $\rho(T) \ne 0$.
However, the present form is still possible
because the phase becomes insignificant 
if $\rho(T) \ne 0$ in which case the kernel vanishes identically.

The remarkable fact in the critical case 
is that one can express the transition kernel (\ckera)
in terms of the stretching factor $k(\lambda)$
and the Morse index $m(\lambda)$ of the potential $\lambda(t)$.
To see this, let us first write the kernel (\ckera) in the form, 
$$
K(b,T;a,0) =  R(T)\, \delta ( \rho(T) )\, e^{ i \Theta (T) },
\eqn\modulpath
$$
with
$$
R(T) =  \sqrt{ 2 \pi }\, | {\cal N} |
        \left[ \prod_{n \ne m} | E_n | \right]^{ -\frac12 }
        | \dot{u}_m (T) ) |^{ -1 }.
\eqn\defR
$$
Then, the unitarity relation,
$$
        \int db\, K^* (b,T;c,0) \, K (b,T;a,0) = 
        \int db\, \langle c | \, 
        \widehat U^\dagger(T,0)\,  | b \rangle
        \langle b | \, \widehat U(T,0)\,  | a \rangle
        = \delta (a-c),
\eqn\unitarity
$$
determines the modulus $R(T)$ to be
$$
R(T)      = \sqrt{\vert k(\lambda)\vert}.
\eqn\squaredR
$$
On the other hand, since the Morse index gives 
the number of negative modes plus one, we have
$$
        \left[ 
        \prod_{n \ne m} E_n
        \right]^{- \frac12}
        =
        \left[ 
        \prod_{n \ne m} | E_n | 
        \right]^{- \frac12}
        e^{- {{i\pi}\over 2} (m( \lambda ) - 1) }.
\eqn\phasepart
$$
Thus the phase part is given by
$$
\Theta (T) = I[x_{\rm cl}] - {\pi\over 2}(m(\lambda) - 1)\ .
\eqn\defS
$$
By combining these we find that the transition kernel
(\ckera) takes the simple form\note{An abstract form
of the formula for the period before the first conjugate 
point has appeared in [\DM].}
$$
        K(b,T;a,0) = 
        \sqrt{|k(\lambda)|}\,
        \delta\bigl( b - k(\lambda) a - s(T) \bigr)\,
        e^{ i I[ x_{\rm cl} ] - {{i\pi}\over 2} m(\lambda) },
\eqn\ckerb
$$
up to an overall constant of unit modulus.  

As application, take, for example, 
the forced harmonic oscillator 
given by $\lambda(t) = \omega^2$ and 
$\mu(t) = - f(t)$.  A special solution satisfying
$s(0) = 0$ is then 
$$
s(t) = {1 \over \omega} \int^t_0 dt'\,
\sin \omega(t - t')\, f(t').
\eqn\spesol
$$
As mentioned earlier for the pure harmonic oscillator case, 
caustics occur at $\omega = {{n\pi}\over T}$ with 
$n = 1, 2, \ldots$, where we have the stretching factor
$k(\lambda) = (-1)^n$ and the Morse index $m(\lambda) = n$.  
The endpoint of the solution
is 
$s(T) = - (-1)^n \int^T_0 dt\, \sin \omega t\, f(t)/\omega$
and the classical action reads
$$
I[ x_{\rm cl} ] = a \int^T_0 dt\, \cos \omega t\, f(t)
- {1\over\omega} \int^T_0 dt \int^t_0 dt'\, 
\cos \omega t\,\sin \omega t'\, f(t)\,f(t')\ .
\eqn\clfho
$$
In particular, for constant $f$ 
the kernel (\ckerb) reduces to
$$
K(b,T;a,0) = \delta
\Bigl( 
b - (-1)^n a - \{ 1 - (-1)^n \} f/\omega^2
\Bigr)\,
        e^{ i( f^2T/2\omega^2 - n\pi/ 2) }.
\eqn\kernfho
$$
This agrees with the results obtained earlier
[\Souriau, \Horv, \C] 
by other indirect means\note{The result contained in [\C] 
for $f(t) \ne 0$ is marred by an error in the phase.}
for the (forced) harmonic oscillator.

It is possible to combine 
the two kernel formulae, (\kernelnonc) and (\ckera), 
into one expression.  For this, we just introduce an
infinitesimal $\epsilon > 0$ to replace $E_n$ with
$E_n + i \epsilon$ in (\calculus), and perform the
integrations over $a_n$ (which are now all Gaussian
for any $\lambda$) to get
$$
        K(b,T;a,0)  
        = {\cal N}
        \left[ \prod_n (E_n + i \epsilon) \right]^{ -\frac12 }\,
        e^{ i \Phi(b, a; \lambda) }\ ,
\eqn\genkernel
$$
where
$$
\eqalign{
\Phi(b, a; \lambda) 
&:= I[ \bar{x}_{\rm cl} ] + I[ \rho ]\vert_{\mu = 0} + 
\dot{\bar{x}}_{\rm cl}(T) \rho(T) \cr
&\qquad - \frac{1}{2} \sum_n \frac{1}{ (E_n  + i \epsilon)}
                \left\{
                        \rho(T)\, \dot{u}_n (T)
        + (E_n + i \epsilon) \int_0^T dt \, \rho(t)\, u_n(t)
                \right\}^2\ .
}
\eqn\genphase
$$
Now for $\lambda$ non-critical, 
the kernel (\genkernel) reduces to
(\kernelnonc) in the limit $\epsilon \rightarrow 0$ by construction.
(This can also be confirmed explicitly; see Appendix.) 
But it is also easy to see that, for $\lambda$ critical,
it reduces to (\ckera) in the limit, since the mode $m$ with
$E_m = 0$ yields the required delta-function on 
account of the identity,
$
\lim_{\epsilon \rightarrow 0} 
\frac{1}{ \sqrt{2 \pi \epsilon} } \,
        e^{ - {x^2}/{2\epsilon}}  = \delta(x)\ .
$
This expression (\genkernel) with (\genphase) will be useful
later in section 4 when we examine the correspondence
hypothesis.

\medskip
\noindent
{\bf 3.2. Quantum effect in the Gaussian slit experiment}

For $\lambda$ non-critical but close to a critical 
$\bar\lambda$,
one expects that there will be a concentration in the
transition amplitude around the focal point 
$b = k(\bar\lambda) a - s(T)$ 
conjugate to a given initial point $a$.
We now analyze how the concentration takes place
quantum mechanically, based on the Gaussian slit  
experiment which is 
laid out in the textbook of Feynman-Hibbs [\FH].

To illustrate our point, let us first consider the harmonic 
oscillator $\lambda = \omega^2$ (with $\mu = 0$).
In the Gaussian slit
experiment we look at the evolution
of a Gaussian wave packet prepared at $t = 0$, and 
for this we put an apparatus 
which emits a particle from the origin $x = 0$
at time $t = - \tau$.  To get a Gaussian
distribution at $t = 0$, we place a \lq Gaussian 
slit' centered at $x = a$ with effective width (variance) 
$\sigma_0$.
The slit 
is realized by means of the Gaussian transmission factor,
$$
G(x - a; \sigma_0) = N\, 
\exp\left\{
-{{(x - a)^2}\over{4\sigma_0^2}}
\right\}.
\eqn\gaussfactor
$$
The wave function $\psi(x, 0)$ of the particle at $t = 0$
is then furnished by the product of 
the free particle kernel\note{
To render the quantum effect manifest, 
we keep $\hbar$ throughout this subsection.}
$
K_0(x, 0; 0, -\tau) 
= \sqrt{1/(2\pi i \hbar \tau)}\, e^{i x^2/2\hbar\tau }
$
and the transmission factor.  From the 
normalization condition
$\int dx \, \vert \psi(x, 0) \vert^2 = 1$ we determine 
the constant $N$ and obtain
$$
\psi(x, 0) =  G(x-a; \sigma_0)\,  K_0(x, 0; 0, -\tau)
= {1\over{(2\pi\sigma_0^2)^{1\over 4}}}\, 
\exp\left\{
-{{(x - a)^2}\over{4\sigma_0^2}} 
+ {{i x^2 }\over{2\hbar \tau}}
\right\}.
\eqn\initialwave
$$
Note that the initial state (\initialwave) has
the average momentum
$$
p = \int dx\, \psi^*(x, 0)
\left(-i\hbar {d\over{dx}}\right) \psi(x, 0) 
= {a \over \tau}.
\eqn\avmon
$$

At $t = T$, the wave function is given by
$$
\eqalign
\psi(y, T) = \int dx\, K(y, T; x, 0)\, \psi(x, 0),
\eqn\finalwave
$$
with the kernel for the harmonic oscillator [\Souriau],
$$
K(y, T; x, 0) 
= \left(
{\omega\over{2\pi i \hbar \vert \sin \omega T \vert }}
\right)^{1\over 2}
\exp\left(
{{i\omega}\over{2\hbar \sin \omega T}} 
\left\{ (x^2 + y^2)\cos \omega T - 2xy 
\right\}
\right)\, 
e^{ - {{i \pi}\over 2} {\rm Int}
\left[ {{\omega T}\over{\pi}} \right] },
\eqn\hokernel
$$
where Int$[x]$ stands for the greatest integer below $x$.
Together with (\initialwave) the integration in (\finalwave)
is Gaussian and hence we get
$$
\psi(y, T) = 
{1\over{(2\pi\sigma^2(T))^{1\over 4}}} 
\exp\left\{
-{
{\left( y - x_{\rm cl}(T)  
\right)^2
}
\over{4\sigma^2(T)}
} 
+ i\, ({\rm phase}) 
\right\}.
\eqn\finalwf
$$
Thus, as is well-known, the Gaussian distribution retains
its shape for the harmonic oscillator, in which the center moves
along the classical path having the initial momentum $p$,
$$
x_{\rm cl}(T)
= a\cos \omega T + {p \over \omega} \sin \omega T,
\eqn\clpath
$$
while the variance pulsates as 
$$
\sigma(T) = \sigma_0 
\left\{
\left({{x_{\rm cl}(T)}\over{a}}\right)^2 
+ 
\left({{\hbar\sin\omega T}\over{2\sigma_0^2\omega}}
\right)^2 
\right\}^{1\over 2}.
\eqn\variance
$$
Clearly, the second term in (\variance)
represents the quantum effect
whereas the first term 
is just the classical variance,
since the variation in the initial position
$\Delta x_{\rm cl}(0) = \sigma_0$ around the center 
$x_{\rm cl}(0) = a$ for the classical path results in 
the final variation
$\Delta x_{\rm cl}(T) = \sigma_0 \vert x_{\rm cl}(T)\vert/a$.
(Notice that the initial momentum $p$ is linearly dependent
on the initial position where the path goes through.)
Thus, as one expects, the quantum effect always enhances the spread
of the Gaussian distribution.

Viewed as a function of the initial variance $\sigma_0$,
the final variance $\sigma(T)$ in (\variance) attains
its minimum, {\it i.e.}, the highest concentration 
of intensity,
$$
\sigma_{\rm min}(T) =
\left\vert
{{\hbar x_{\rm cl}(T) \sin\omega T}\over{a \omega}}
\right\vert^{1\over 2},
\eqn\minivar
$$
at 
$$
\sigma_0 = \left\vert
{{\hbar a \sin\omega T}\over{2 \omega x_{\rm cl}(T)}}
\right\vert^{1\over 2},
\eqn\inimini
$$
which is precisely the point where the quantum effect
matches the classical variance.
The infinite concentration $\sigma_{\rm min}(T) = 0$
takes place at 
$\omega = n\pi/T$, or at
$p = - a \omega/\tan \omega T$.  
The former case is the expected caustics, 
where the variance (\variance)
becomes purely classical and reproduces 
the initial value $\sigma(T) = \sigma_0$.  Thus
the infinite concentration is obtained if we let 
$\sigma_0 \rightarrow 0$.  By contrast,
in the latter case 
the variance becomes purely quantum mechanical, 
as all paths passing through the slit   
coalesce
toward the origin $x = 0$ at $t = T$.
In fact, this case is again caustics occurring during 
the combined period $[-\tau, T]$ 
under the potential $\lambda(t)$ which vanishes
for $t < 0$.  
Note that in this case
the infinite concentration at the quantum level 
is achieved by letting $\sigma_0 \rightarrow \infty$. 
In practice, however, 
both of these concentrations in intensity 
are unstable, since a small fluctuation in 
the parameters $\omega$ 
or $\tau$ will generally bring the variance to a large value,
as can be seen in Fig.2.
Nevertheless, one could achieve a very high intensity
by adjusting the initial variance along with the parameters
according to (\inimini).
\topinsert
\vskip 1.4cm
\let\picnaturalsize=N
\def\picsize{7.5cm}
\def\picfilename{fig2a.epsf}
\input epsf
\epsfxsize \picsize 
\hskip 2.8cm\epsfbox{\picfilename}
\vskip 1.5cm
\let\picnaturalsize=N
\def\picsize{7.5cm}
\def\picfilename{fig2b.epsf}
\input epsf
\epsfxsize \picsize 
\hskip 2.8cm\epsfbox{\picfilename}
\vskip 0.5cm
\abstract{%
{\bf Figure 2.} 
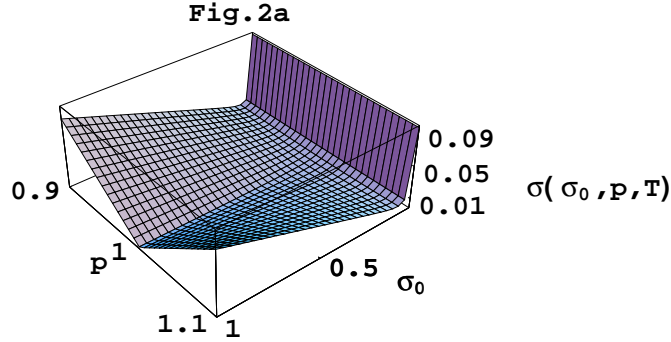
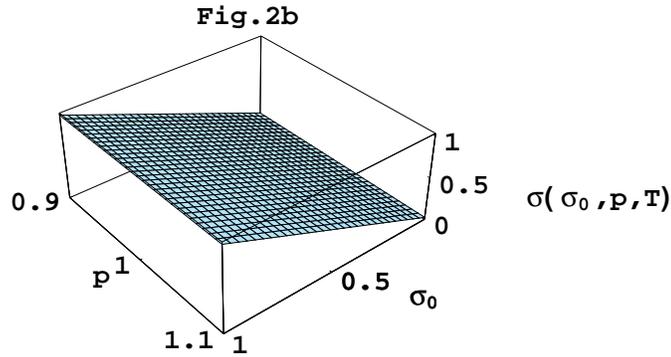
The final variance $\sigma(T) = \sigma(\sigma_0, p, T)$ 
as a function of the initial
variance $\sigma_0$ and the momentum $p$.  Fig.2a shows
a generic case given by the parameters $a = 0.3$, $\omega = 2.5$ 
and $T = 1.0$ (we set $\hbar = 0.004$ here).  
The focal points due to the caustics associated with the period
$[-\tau, T]$  
appear at the bottom of the valley at $p = 1.0$.  
The minimum value $\sigma_{\rm min}(T)$ at the bottom tends to
vanish as $\sigma_0 \rightarrow \infty$ but it is unstable under the 
fluctuation of $p$. 
When $\omega$ reaches the first critical value $\omega = \pi$
associated with the period $[0, T]$, the situation changes 
drastically to the one shown in Fig.2b. 
This time the minimum $\sigma_{\rm min}(T) = 0$ attained at 
$\sigma_0 = 0$ is unstable under the fluctuation of 
$\omega$.     
}
\endinsert

An important quantity for characterizing the concentration
is the susceptibility of the variance against initial momentum 
fluctuations.
Exposing the momentum dependence of
the variance explicit $\sigma(T) = \sigma(p, T)$,
we use the following normalized quantity for
the susceptibility,
$$
S(p, T) :=
{a \over {\sigma_0}} {{\partial}\over{\partial p}}
\sigma(p, T).
\eqn\qjacobi
$$
For the present harmonic oscillator
(\variance) we find
$$
S(p, T) = \left\vert {{\sin \omega T}\over{\omega}}
\right\vert
\left\{
1 + 
\left(
{{\hbar \sin \omega T}\over{2 \sigma_0^2 \omega x_{\rm cl}(T)}}
\right)^2
\right\}^{-{1\over 2}}.
\eqn\qjharmonic
$$
In the classical limit 
$\hbar \rightarrow 0$, the susceptibility 
reduces to the (absolute value of the)
Jacobi field, which in this case becomes 
$J(p, T) = {\sin \omega T}/{\omega}$.  
In the two cases mentioned above 
in which an infinite concentration can in principle 
be possible,
the susceptibility vanishes $S(p, T) = 0$ and hence 
the Gaussian wave packet becomes free from
momentum fluctuations.
What is interesting in this result (\qjharmonic), however, 
is that the quantum effect suppresses the susceptibility of
the variation against initial momentum fluctuations.

We now show that these features are universal 
for a generic potential $\lambda$ if $\mu = 0$.
For this we need to note two crucial points.
First, for systems with
quadratic Lagrangians the semiclassical (WKB) 
approximation for the kernel gives the exact result,
$$
K(y, T; x, 0) = \left(
{{i}\over{2\pi\hbar}} 
\left\vert
{{\partial^2 I_{\rm cl}}
\over{\partial y \partial x}}
\right\vert
\right)^{1\over 2}\, 
e^{
{i \over {\hbar}} I_{\rm cl} - {{ i\pi}\over 2} m(\lambda)  
}\ ,
\eqn\wkb
$$
where $I_{\rm cl} = I_{\rm cl}(y, T; x, 0) := I[x_{\rm cl}]$
is the action evaluated for the classical
path $x_{\rm cl}(t)$ connecting the endpoints, 
$x_{\rm cl}(0) = x$ and $x_{\rm cl}(T) = y$.  

Second, if $\mu = 0$ 
the classical action $I_{\rm cl}(y, T; x, 0)$ is
a quadratic polynomial homogeneous in $x$ and $y$,
because we have 
$$
I_{\rm cl}(cy, T; cx, 0) = I[c x_{\rm cl}]
= c^2 I[x_{\rm cl}] = c^2 I_{\rm cl}(y, T; x, 0)\ ,
\eqn\homog
$$ 
for any constant $c$.  
Thus, with $A$, $B$ and $C$ being some
functions of $T$, we may write
$$
I_{\rm cl}(y, T; x, 0) = A x^2 + B xy + C y^2\ .
\eqn\claction
$$
Then it is straightforward to see that 
the Gaussian wave packet
(\initialwave) evolves into  
the wave function (\finalwave) at time $t = T$ 
in exactly the same form (\finalwf) with
$$
x_{\rm cl}(T) = - {1 \over B} (2 a A + p),
\eqn\gencl
$$
and
$$
\sigma(T) = \sigma_0 
\left\{
\left({{x_{\rm cl}(T)}\over{a}}\right)^2 
+ 
\left({{\hbar}\over{2\sigma_0^2 B}}
\right)^2 
\right\}^{1\over 2}.
\eqn\genvar
$$
The variance attains its minimum,
$$
\sigma_{\rm min}(T) = 
\left\vert
{{\hbar x_{\rm cl}(T)}\over{a B}}
\right\vert^{1\over 2}
\qquad
\hbox{at}
\quad 
\sigma_0 = \left\vert
{{\hbar a }\over{2 B x_{\rm cl}(T)}}
\right\vert^{1\over 2}.
\eqn\geninimini
$$

The susceptibility 
of the variance (\qjacobi) then reads
$$
S(p, T) = \left\vert J(p, T) \right\vert
\left\{
1 + 
\left(
{{\hbar a}\over{2 \sigma_0^2 B x_{\rm cl}(T)}}
\right)^2
\right\}^{-{1\over 2}},
\eqn\qjharmonic
$$
where $J(p, T) = - \left(\partial^2 I_{\rm cl}
/\partial y \partial x \right)^{-1} = - 1/B$ 
is the Jacobi field for the classical action (\claction).
The generic behavior of the susceptibility 
(\qjharmonic) with respect to $p$
is analogous to 
what we have seen in the harmonic oscillator case.

The above features of the Gaussian wave packet 
persist even for $\mu \ne 0$, 
because then 
the classical action $I_{\rm cl}(y, T; x, 0)$ acquires
only linear and constant terms in addition to the
quadratic terms in (\claction) as can be explicitly 
seen from the form
of the general classical solution (\helpa). 
Accordingly, the time evolution
by the integral (\finalwave) is essentially unchanged.
Semiclassically, this may also be the case for more general systems, 
not only for those with 
quadratic Lagrangians we considered, in view of
the earlier study [\H] which suggests that these features
are a norm for a generic system in the limit $\hbar \rightarrow 0$. 

\ve

\secno=4 \meqno=1


\centerline
{\bf 4. The Role of Instantons in the Quantum Mechanical Rotor}

\medskip

In the previous section we obtained the path-integral kernel 
for the quadratic action with a generic time-dependent
potential function 
$\lambda(t)$.  The aim of 
this section is to apply it to the problem
of quantum mechanical rotor on the circle $S^1$ 
and thereby examine Jevicki's 
correspondence hypothesis [\Jevicki] 
mentioned in the Introduction.
We first furnish a general scheme
to compute the path-integral kernel for the rotor based on
the formula we just obtained.  This will 
strengthen the basis on which Jevicki's arguments for 
the hypothesis stand and highlight 
their loose ends at the same time.
Our close examination will show that neither of the two
crucial assumptions adopted in his arguments
can be entirely justified for generic transition amplitudes.

\medskip

\noindent
{\bf 4.1. Examination of the correspondence hypothesis}

The system taken up by Jevicki to illustrate the
correspondence (\hypo) between stationary points and
simple poles is the quantum mechanical  
rotor, namely, a free particle on the circle $S^1$.  
Embedded in the two dimensional plane $\Real^2$, 
the system acquires a harmonic potential and develops
a singularity in the transition kernel, and this  
has been used as a prototype to argue that 
the presence of instantons 
can be seen as simple poles (rather than stationary points)
in the $\Complex P^N$ nonlinear sigma model.
A salient feature
of the $S^1$ system is that 
its non-trivial topology 
--- $S^1$ is multiply-connected: $\pi_1(S^1) = \Zed$ ---
admits multi-winding configurations 
called `instanton' solutions labelled by the 
winding number $n \in \Zed$.
Here, our concern lies in
the question whether the instantons can 
also be seen explicitly even if we 
embed the $S^1$ in the topologically trivial
plane $\Real^2$ using
the Cartesian coordinates.
Before enforcing this `Cartesian point of view', 
we wish to recapitulate the known result
of the path-integral on $S^1$ in the `intrinsic point of view',
that is, when the circle is regarded as intrinsic
rather than being embedded in $\Real^2$, where
the instantons arise as sole contributors to the 
transition kernel.

For a free particle on $S^1$ coordinated by the
angle $\varphi \in [0, 2\pi)$ we have the classical action 
$I^{S^1}_0[\varphi] = \int_0^T dt {1\over 2}(\dot \varphi)^2$.
In the path-integral formalism, the transition amplitude 
going from $\alpha$ at $t = 0$ to $\beta$
at $t = T$ where $\alpha$, $\beta \in [0, 2\pi)$ is given by 
$$
K^{S^1}_0( \beta, T; \alpha, 0 ) 
= \int^{\varphi(T) = \beta}_{\varphi(0) = \alpha} 
{\cal D}\varphi \, e^{iI^{S^1}_0[\varphi]}\ .
\eqn\formal
$$
This formal expression must be given a meaning
such that it accommodates multi-winding paths 
allowed for the transition since $S^1$
is multiply-connected.
The conventional method [\Schulman] 
to deal with such paths is that,
instead of working on $S^1$, one considers
the corresponding free particle system on the 
covering space of the circle, {\it i.e.}, the line
$\Real$ governed by the action
$I_0[x] = \int_0^T dt {1\over 2}(\dot x)^2$, 
which is invariant under the translation by $2\pi$.
Namely, to the amplitude on $S^1$ consisting of
paths of winding number $n$, one assigns the free particle 
amplitude
$K_0 ( \beta + 2 \pi n, T; \alpha, 0 )$ on the line $\Real$,
and then sums it up over the integers $n \in \Zed$
with a weight factor $w(n)$,
$$
K^{S^1}_0 ( \beta, T; \alpha, 0 ) =
  \sum_{ n = - \infty }^{ \infty } \, w(n) \,
        K_0 ( \beta + 2 \pi n, T; \alpha, 0 ).
\eqn\multi
$$

In short, this amounts to regarding 
$S^1$ as the coset space ${\Real}/{\Zed}$
by identifying those points on the line ${\Real}$ which differ
by $2\pi \times \hbox{integer}$.  
The weight factor $w(n)$ appearing in (\multi) signals 
the ambiguity that can arise due to the multiply-connectedness
of the space.  It is furnished by the unitary representation 
$w(n) = e^{in\theta}$ of the group $\Zed$,
where $\theta \in [0, 2\pi)$ is the angle parameter
specifying the representation and, hence, the ambiguity.\note{
For a general discussion of the path-integral
quantization on a coset space $G/H$, see [\TT].
} 
In what follows, however, we put $\theta = 0$ 
for simplicity.  
Recall that on the
line $\Real$ the kernel for the free particle is
$
K_0 ( b, T; a, 0 ) = \int^{x(T) = b}_{x(0) = a} {\cal D}x\,
e^{I_0[x]} = \sqrt{1/(2\pi i T)}\, e^{iI_0[x_{\rm cl}]}
$
with 
$I_0[x_{\rm cl}] = (b - a)^2/2T$.
In our case, the classical
solution satisfying the required boundary conditions on $S^1$
possessing winding number $n$, the $n$-instanton solution, 
is given by
$$
x_{\rm cl}^{(n)}(t) = 
(\beta - \alpha + 2n\pi ){t\over T} + \alpha\ .
\eqn\instanton
$$
Accordingly, 
the kernel (\multi) on the circle $S^1$ becomes
$$
K^{S^1}_0 ( \beta, T; \alpha, 0) = 
\sqrt{{1\over{2\pi i T}}}
  \sum_{ n = - \infty }^{ \infty } 
e^{i {(\beta - \alpha + 2n\pi )^2}/{2T} }\ ,
\eqn\ampsone
$$
which is just a sum of instanton contributions.

In the Cartesian coordinate point of view,
on the other hand,
one puts the particle on the two dimensional plane coordinated
by ${\bf x} = (x, y) \in \Real^2$ and imposes the constraint
$\vert{\bf x}\vert^2 = 1$.  The constraint can be
implemented by introducing the Lagrange multiplier $\lambda(t)$
and writing (\formal) as
$$
\eqalign{
K^{S^1}_0( \beta, T; \alpha, 0 )
&= \int_{{\bf x}(0) 
   = (\cos \alpha, \sin \alpha)}^{{\bf x}(T) 
   = (\cos \beta, \sin \beta)} 
   {\cal D}\lambda\, {\cal D}{\bf x}\,
   \exp \Bigl[{i\over 2}\int^T_0 dt\, \bigl( 
   \vert \dot {\bf x}\vert^2 
   - \lambda (\vert {\bf x} \vert^2 -1) \bigr) \Bigr] \cr
&= \int {\cal D}\lambda\, 
   e^{{i\over 2}\int^T_0 dt\, \lambda} \, 
   \int_{{x}(0) = \cos \alpha}^{{x}(T) = \cos \beta} {\cal D}{x}\,
   e^{i I[x]} \,  
   \int_{{y}(0) = \sin \alpha}^{{y}(T) = \sin \beta} {\cal D}{y}\,
   e^{i I[y]}\ ,
}
\eqn\sopi
$$
where $I[x]$ (and similarly $I[y]$)
is the quadratic action with time-dependent 
harmonic potential $\lambda(t)$ under zero force $\mu(t) = 0$.
In words, the kernel is given by 
the product of two kernels on the line sharing the potential
$\lambda(t)$, one of which describing the transition
>from $\cos \alpha$ to $\cos \beta$
and the other from $\sin \alpha$ to $\sin \beta$,
averaged over $\lambda(t)$ with the weight factor
$e^{{i\over 2}\int^T_0 dt\, \lambda}$,
$$
K^{S^1}_0( \beta, T; \alpha, 0 )
= \int {\cal D}\lambda\, 
K(\cos \beta, T; \cos \alpha, 0)\, 
K(\sin \beta, T; \sin \alpha, 0)\,
e^{{i\over 2}\int^T_0 dt\, \lambda} \ .
\eqn\sokernel
$$
Using the expression (\genkernel) for $K(b, T; a, 0)$ 
obtained in section 3, the kernel (\sokernel) becomes
$$
K^{S^1}_0( \beta, T; \alpha, 0 )
= \int {\cal D}\lambda\,
\left[ \prod_n (E_n + i\epsilon) \right]^{-1} \,
e^{i\Phi(\cos \beta, \cos \alpha; \lambda) +
   i\Phi(\sin \beta, \sin \alpha; \lambda)  
+ {i\over 2} \int^T_0 dt\, \lambda}\ .
\eqn\pik
$$

Now the generic formula for the 
phase (\genphase) implies that the phase 
$\Phi(b, a ;\lambda)$ reduces simply to the action
$I_{\rm cl}(b, T; a, 0) = I[x_{\rm cl}]$ 
except for the cases where the potential $\lambda$ 
is critical and $b$ is not conjugate to $a$, 
{\it i.e.}, $b \ne k(\lambda)a$.  
However, these exceptional cases 
may be neglected in the integration over $\lambda$ 
on the grounds that
they form a set of measure zero 
in the entire space of potentials $\lambda$ and that 
the kernel $K(b, T; a, 0)$ vanishes at those $\lambda$ and 
hence cannot contribute to the integral (\sokernel). 
For this reason
we replace the phase factor in the integration
by the classical action with the prescription,
$$
\Phi(b, a;\lambda) \rightarrow 
I^{(\xi)}_{\rm cl}(b, T; a, 0)
:= (1 + i\,{\hbox{sign}(I)}\xi)\,I_{\rm cl}(b, T; a, 0)\ .
\eqn\prescription
$$
The prefactor $(1 + i\,{\hbox{sign}(I)}\xi)$, where  
$\hbox{sign}(I) := I/\vert I \vert$ gives the sign of
the action with $\xi > 0$ being infinitesimal, is attached 
in order to guarantee that 
$e^{i (1 + i\,{\rm sign}(I) \xi) I}$ vanishes for those critical
potentials with $b \ne k(\lambda)a$ for which
the action diverges (see Lemma 9).

To proceed further and make contact with Jevicki's
argument, let us 
consider the separation of the zero mode part $\lambda_0$ 
>from the potential,
$$
\lambda(t) = \lambda'(t) + \lambda_0, \qquad \hbox{with} \quad 
\int_0^T dt\, \lambda'(t) = 0, \quad
\lambda_0 = \frac{1}{T} \int_0^T dt\, \lambda(t)\ .
\eqn\separation
$$  
As we did for the potential $\lambda(t)$ in (\eigen) and (\usn),
we can also consider a complete set of orthonormal 
eigenfunctions for the non-zero mode part $\lambda'(t)$.
Let the eigenvalues of those eigenfunctions be
$\varepsilon_n$.  Obviously, both $\lambda(t)$ and $\lambda'(t)$
share the same set of eigenfunctions with 
eigenvalues related by
$$
E_n = \varepsilon_n - \lambda_0.
\eqn\eigenrel
$$
Accordingly, a potential $\lambda(t)$ is critical 
if $\lambda_0 = \varepsilon_n(\lambda')$ for some $n$.

Substituting (\eigenrel) in (\pik), we find
$$
\eqalign{
K^{S^1}_0( \beta, T; \alpha, 0 )
&= \int {\cal D}\lambda' d\lambda_0\,
\left[ 
\prod_n (\varepsilon_n - \lambda_0 + i\epsilon) 
\right]^{-1} \cr
& \times \exp 
\left\{
iI^{(\xi)}_{\rm cl}(\cos \beta, T;  \cos \alpha, 0) +
   iI^{(\xi)}_{\rm cl}( \sin \beta, T; \sin \alpha, 0)
   + {i\over 2} \lambda_0 T
\right\}\ .
}
\eqn\finalpik
$$
Consider now the integration over the zero mode 
$\lambda_0$.  Observe that 
the integrand diverges (in the limit $\epsilon \rightarrow 0$)
for critical $\lambda$, because 
of the factor 
$[\prod_n (\varepsilon_n - \lambda_0 + i\epsilon)]^{-1}$.
Recall that the additional 
divergence that could arise
in the actions $I_{\rm cl}(\cos \beta, T;  \cos \alpha, 0)$ and  
$I_{\rm cl}( \sin \beta, T; \sin \alpha, 0)$
is taken care of by the prefactor $(1 + i\,{\rm sign}(I)\xi)$.
Thus we are naturally led to evaluate the integration
over the zero mode by collecting the values at the
critical potentials which admit the endpoint $b$
to be conjugate to $a$
in both $x$ and $y$-direction.  
This will be carried out by
regarding the $\lambda_0$ integration 
as part of the contour integration in the 
complex $\lambda_0$ plane as shown 
in Fig.3.
Since each critical potential which satisfies 
the above condition provides a simple pole in the integrand
in (\pik), and since the contribution 
>from the semicircle of the contour 
vanishes as the radius becomes large
due to the weight factor 
$e^{{i\over 2}\lambda_0 T}$,
the $\lambda_0$ integration boils down to
the residue calculus at those simple poles.
That the $\lambda_0$ integration in (\pik) 
can be carried out by 
the contour integration in the way described above,
ignoring the contributions from those critical potentials
which do not meet the 
condition $b = k(\lambda)a$,
constitutes our {\it first assumption}.
\topinsert
\vskip 1.4cm
\let\picnaturalsize=N
\def\picsize{7.5cm}
\def\picfilename{fig3.epsf}
\input epsf
\epsfxsize \picsize 
\hskip 2.8cm\epsfbox{\picfilename}
\vskip 0.5cm
\abstract{%
{\bf Figure 3.} 
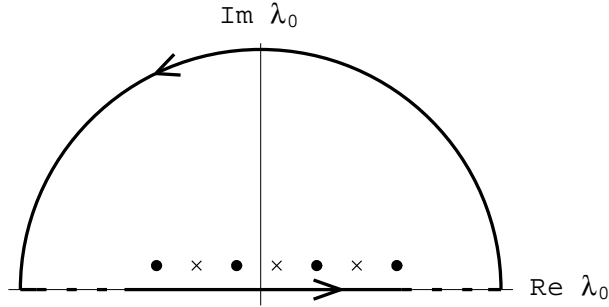
The integration contour in the complex $\lambda_0$-plane.
Among the potentially singular points
$\lambda_0 = \varepsilon_n + i\epsilon$,
those critical potentials which meet the 
condition stated in the text
give rise to simple poles 
shown by \lq $\bullet$'.  By contrast, those which do not 
meet the condition, shown here by \lq $\times$', 
will not create actual singularities for 
contributing to the integration.
}
\endinsert

Unfortunately, this assumption cannot be true in general.
Indeed, the condition for a critical potential $\lambda(t)$ to 
create just a simple pole (not other type of potential singularities)
is to have the 
stretching factor $k(\lambda)$ which matches the boundary conditions
in both $x$ and $y$-direction simultaneously,
$$
k(\lambda) = {{\cos \beta}\over {\cos \alpha}}
\qquad \hbox{and} \qquad
k(\lambda) = {{\sin \beta}\over {\sin \alpha}}\ ,
\eqn\charac
$$
but, obviously, these are 
incompatible in general.  It then follows 
>from the absence of simple poles that 
the transition amplitude (\pik) on the circle must
vanish for generic boundary conditions, a result in 
contradistinction with (\ampsone).
Having no simple poles that could
correspond to the stationary points in the intrinsic viewpoint,
we find that 
the correspondence hypothesis (\hypo) cannot be true in general.
It is also worth mentioning that in the
intrinsic viewpoint the
stationary points 
contributing to (\ampsone) do not in general 
correspond to the stationary points
appearing under the critical 
potentials in the Cartesian viewpoint.

The condition (\charac)
becomes compatible if $\tan \alpha = \tan \beta$, 
that is, if the two endpoints are the same 
or diametrically opposite with each other. 
To examine whether
it is possible to sustain the correspondence hypothesis
if restricted to this specific case, let us take 
$\alpha = \beta = 0$ from now on.
Then we observe that, in this case, we have
$I_{\rm cl}(0, T; 0, 0) = 0$ for the classical action
in $y$-direction, and 
the condition is fulfilled if $k(\lambda) = 1$.
We thus arrive at
$$
K^{S^1}_0( 0, T; 0, 0 )
= F(T) 
\int {\cal D}\lambda'\, \sum_n e^{i \Phi_n(\lambda')}\ ,
\eqn\pif
$$
with the total phase
$$
\Phi_n(\lambda')  
:= \left(
I^{(\xi)}_{\rm cl}(1, T; 1, 0)
- m(\lambda)\pi  + {1\over 2} \lambda_0 T 
\right)\biggr\vert_{\lambda_0 = \varepsilon_n}  \ ,
\eqn\totphase
$$
where we formally included all the contributions from
critical potentials using (\ckerb) in view of the fact that
those which correspond to $\lambda$ with $k(\lambda) \ne 1$
will drop out due to the prefactor attached in (\prescription).   
In (\pif), the real function $F(T)$ accounts for
the possible overall factor that arises
in the change of integration measure and the
residue computation, and $m(\lambda)$
is the index of the critical potential $\lambda(t)$. 
An important point to note is that, 
for any critical potential with $k(\lambda) = 1$ 
that can contribute to the kernel (\pif),
the index $m(\lambda)$ is always even on account of
the alternate nature of the sign of 
the stretching factors $k(\lambda)$ (see Lemma 11).
Thus we may drop the index term $m(\lambda)\pi$
in the total phase $\Phi_n(\lambda')$ 
without affecting the kernel (\pif).

According to Jevicki [\Jevicki],
we now pose the {\it second assumption}:
the total phase factor $\Phi_n(\lambda')$ 
in (\pif) is independent of $\lambda'(t)$.
This is a rather drastic assumption, allowing 
us to set $\lambda'(t) = 0$ and getting 
the eigenfunctions and the eigenvalues, 
$$
u_n(t) = \sqrt{{2 \over T}} 
\sin \left({{n\pi t}\over T}\right)\ ,
\qquad
\varepsilon_n = \left( {{n\pi}\over T} \right)^2\ ,
\eqn\efho
$$
for $n = 1$, 2, 3, $\ldots$.  
We then find that the critical potentials
arise at $\lambda(t) = \lambda_0 = \varepsilon_n$, 
and they have the stretching factor 
$k(\lambda = \varepsilon_n) = (-1)^n$.
Thus, among them those possessing $k(\lambda) = 1$
occur at even integers $n = 2l$.  Since
we have $I_{\rm cl}(1, T; 1, 0) =0$
for the classical solution
in the (critical) harmonic oscillator, 
and since the index is given by
$m(\lambda = \varepsilon_{2l}) = 2l$, 
the total phase becomes
$$
\Phi_{2l}(0) = - 2l\pi + {{2 (l \pi)^2}\over{T}}\ .
\eqn\tphase
$$
Consequently, we obtain
$$
K^{S^1}_0( 0, T; 0, 0 ) = {1\over 2}F(T) \sum_{l \ne 0}
e^{i {{2 (l \pi)^2}/{T}}}\ .
\eqn\finalpi
$$
Thus, if we choose $F(T) = 2 \sqrt{{1 \over{2 \pi i T}}}$,
the result (\finalpi) coincides with (\ampsone), 
up to the contribution from $l = 0$.  

The foregoing argument is 
(a rigorous version of) the one used in [\Jevicki]
which asserts that, even in the Cartesian point of view,
one can see the instantons  
contributing to the kernel by the form
of simple poles, rather than stationary points, 
in the integrand.  
In fact, for the boundary condition $\alpha = \beta = 0$
the classical motions allowed under 
the critical potentials sitting at the simple
poles correspond precisely 
to the instanton solutions (\instanton)
which are the stationary
points in the intrinsic viewpoint.
Thus we find that the correspondence 
hypothesis (\hypo) holds if one 
restricts oneself to the specific transition process and   
ignores the lacking piece $l = 0$ in (\finalpi)
which, if existed,  
would correspond to the trivial ($0$-instanton) 
solution $x^{(0)}_{\rm cl}(t) = 0$
in the intrinsic viewpoint.

\medskip

\noindent
{\bf 4.2. Validity of the second assumption: an example}

Although we have seen that the
first assumption cannot be sustained for generic
transition processes, at the end of the computation
the kernel (\pik) must recover 
the integers ({\it i.e.}, instanton numbers)
appearing in (\ampsone) for any $\alpha$ and $\beta$.
It is a matter of fact that the classical
motions expressed by the
instantons $x_{\rm cl}^{(n)}(t)$ in (\instanton) arise at
$\lambda(t) = \hbox{constant}$, even though they
do not necessarily appear as simple poles for generic
$\alpha$ and $\beta$.  
In this respect the second assumption
does lead to the correct result in effect, and one is 
curious whether there is an {\it a priori} reason for this.
(Note that the two assumptions are independent of 
each other.) 
In the rest of this section 
we shall investigate this issue 
by looking at a simple but nontrivial example.

To this end, we return to the special case 
$\alpha = \beta = 0$ where the assertion of 
the first assumption is valid 
(up to the contribution at $l = 0$).  
The second assumption now turns out to be
that, modulo $2\pi$, 
the total phase $\Phi_n(\lambda')$ in (\totphase) is
given by the phase $\Phi_n(\lambda' = 0)$ in (\tphase)
with $n = 2l$.  
To make the 
issue more transparent, let us fix some $\lambda'(t)$
and consider
the family of potentials $\lambda(t)$ 
which differ by constants
$\lambda(t) = \lambda'(t) + \lambda_0$
as in the form (\separation).  
Within this family, the critical potentials which arise at
$\lambda_0 = \varepsilon_n$ for $n \in \Zed$ form a class and
can be labelled by the integers $n$.
In particular, those that have $k(\lambda) = 1$ 
form a subclass in the class and may further be labelled by 
another set of integers $l \in \Zed$.
The second assumption can then be justified if 
the total phase $\Phi_n$ admits the form 
(\tphase) modulo $2\pi$ with the summation $l$
of $n = 2l$ being performed over the critical 
potentials in the subclass.     

To see how this works, take the potential
$$
\lambda'(t) = c \; \delta (t - {t_0}) - {c \over T}\ ,
\eqn\dpot
$$
where $c$ is a constant and the term
$- {c \over T}$ is inserted to ensure $\int dt \lambda' = 0$.
This is a perturbation of the trivial case $\lambda'(t) = 0$ 
in the sense that (\dpot) defines a one-parameter family
of potentials containing the trivial case $c = 0$.
The other parameter $t_0 \in (0, T)$
specifies the moment of impact inflicted on the particle 
and may be determined by requiring that $\lambda(t)$ admits
cases where $k(\lambda) = 1$.  From Lemma 6 we know that 
this requirement is met if there exists a classical solution
$u(t)$ of the equation of motion
vanishing at the endpoints $u(0) = u(T) = 0$  
and further has the same velocity 
$\dot u(0) = \dot u(T)$.  Below we shall seek for 
conditions under which such a 
solution exists, setting 
the initial velocity $\dot u(0) = 1$
for definiteness.

{}For this purpose, we first divide the interval $[0, T]$ into
$[0, t_0]$ and $[t_0, T]$, and call them interval (I) and (II).
Then, in each of the intervals, the solution which respects the
boundary conditions takes, respectively, the form
$$
u_{\rm I}(t) = {1 \over \omega} \sin (\omega t)\ , 
\qquad
u_{\rm II}(t) = -{1 \over \omega} \sin (\omega (T - t))\ .
\eqn\solc
$$
The continuity condition
$u_{\rm I}(t_0) = u_{\rm II}(t_0)$ at $t = t_0$ gives
$$
{1 \over \omega} \sin (\omega {t_0})
= -{1 \over \omega} \sin (\omega (T - {t_0}))\ .
\eqn\condone
$$
The velocity acquires a gap at $t = t_0$
prescribed by the delta-function interaction as
$\dot u_{\rm II } (t_0) - \dot u_{\rm I} (t_0)
+ c\; u_{\rm I} (t_0) = 0$,
which amounts to
$$
\cos (\omega (T - {t_0})) - \cos(\omega {t_0})
+ { c \over \omega } \sin (\omega {t_0}) = 0\ .
\eqn\condtwo
$$
The first condition (\condone) is fulfilled if 
$$
\omega (T-{t_0})=-\omega {t_0}+ 2 \pi l \ ,
\qquad \hbox{or} \qquad
\omega (T-{t_0})=\omega {t_0}+ \pi (2 l + 1)\ ,
\eqn\twocases
$$
for $l \in \Zed$.  To each of these cases, 
the other condition (\condtwo) requires that  
(for $c \ne 0$)
$$
\omega= {{2 \pi l }\over T}, \quad c = \hbox{arbitrary}, 
\quad {t_0}={m \over {2 l}} T, 
\eqn\fstcase
$$
where $m$ can be any integer with $ 1 \le m \le 2l-1$, or
$$
\omega={ { (2 l +1) \pi } \over { T-2 {t_0} } }, 
\quad  
c = { {2 \omega \cos(\omega {t_0})} \over {\sin(\omega {t_0})} } \ ,
\quad
t_0 = \hbox{arbitrary}\ .
\eqn\sndcase
$$
Note that the conditions (\sndcase) do not allow for a set
of infinite number of solutions for 
fixed $c$ and $t_0$ (specified by $\lambda'(t)$
that defines the class).  Hence
in the latter case (\sndcase) 
the subclass with $k(\lambda) = 1$ consists 
of finite number (possibly one) of critical potentials.
In the former case (\fstcase), on the other hand, 
if $m = {{lq}\over{p}}$ with $p$ and 
$q$ being coprime positive integers satisfying
${{q}\over{p}} < 2$, then the parameter $t_0$ lies in
the interval $0 \le t_0 \le T$ with the fixed value,
$$
t_0 = {q\over{2p}}T.
\eqn\teezero
$$
Thus the former
case does provide the subclass 
consisting of infinite number of potentials, but
for $m$ to be integers the label $l$ runs as 
$l =  p,\, 2p,\, 3p, \ldots$.
We now examine 
whether our second assumption holds or not here.

To evaluate the total phase explicitly, we 
consider the classical solution $x(t)$ for the above class of  
potentials satisfying the boundary condition,
$$
x(0)=x(T)=1\ , \qquad {\dot x}(0) = 1\ .
\eqn\csbc
$$
(Again, for definiteness we added 
the second condition in (\csbc) because   
the action on caustics does not depend on the initial velocity.)  
As before, in each
of the intervals (I) and (II) the solution
obeying the boundary conditions (\csbc) takes the form,
$$
\eqalign{
{x_{\rm I}}(t)
&={1 \over \omega} \sin(\omega t)+\cos(\omega t)\, \cr
{x_{\rm II}}(t)
&=-{\gamma \over \omega} \sin(\omega(T-t)) + \cos(\omega(T-t))\ ,
}
\eqn\solbc
$$
where $\gamma$ is a constant.
>From the continuity condition 
and the velocity gap equation at $t={t_0}$ in (\teezero)
together with the $\omega$ in
(\fstcase), it is determined to be $\gamma = 1 - c$.
For this solution (\solbc) the classical action becomes
$$
I_{\rm cl}(1, T; 1, 0)
= {1 \over 2} x_{\rm I}(t) \dot x_{\rm I}(t)\Bigr\vert^{t_0}_0
+ {1 \over 2} x_{\rm II}(t)\dot x_{\rm II}(t)\Bigr\vert^T_{t_0}
  - {c \over 2} x_{\rm I}^2(t_0)
= - {c \over 2}\ .
\eqn\valaction
$$
On the other hand, 
the eigenvalues for the critical potentials in the subclass are
$$
\varepsilon_{2l} =  
\left( {{2 \pi l }\over T} \right)^2 + {c \over T}\ ,
\eqn\valeigen
$$
where we put the label $2l$ to match 
the notation (\efho).
Consequently, the total phase (\totphase) turns out to be
$$
\Phi_{2l} = - m(\lambda)\pi + {{2 (l \pi)^2}\over{T}}\ .
\eqn\gtphase
$$
Note that the final expression (\gtphase) is 
independent of the parameter $c$. 
Since the index $m(\lambda)$ is always even as remarked
earlier, we see that, if $p = 1$ for which
the label $l$ takes all integers, the phase factor 
agrees with the original one (\tphase) even if it is
perturbed by the delta-function interaction.
However, for $p > 1$, the total phase
does not reproduce the unperturbed value (\tphase).

We thus find that the assertion of the 
second assumption holds for the family 
of potentials furnished
by (\dpot) with the parameters given by (\fstcase),
if one makes the special choice $p = 1$.
It does not hold, however, for the case (\sndcase), 
where the final phase factor consists of a finite
number of contributions.  In conclusion, there is no
{\it a priori} reason that the second assumption can
lead to the correct transition amplitude, simply because
there are cases where the assumption itself breaks down.
Finally, we point out that the case
$p = 1$ where the assumption does hold
implies $t_0 = T/2$, and this is the
only case in the type of
potentials (\dpot) that admits
all the eigenstates to have a smooth 
transition to the unperturbed ones 
in the limit $c \rightarrow 0$.  
One may speculate that the cases where the non-zero mode 
$\lambda'$-independence is seen, such as the above case
where the cancellation of the $c$-dependence occurs, 
may be found more generally, while
those cases where it is not seen
form a set of measure zero in the total space of 
the non-zero mode of the potential $\lambda'(t)$.

\ve

\secno=5 \meqno=1

     
\centerline{\bf 5. Conclusion and Discussions}
\medskip

In this paper we investigated various aspects of
classical and quantum caustics for quadratic Lagrangians
in one dimension,
and applied the results thereof to examine Jevicki's 
correspondence hypothesis (\hypo) between 
stationary points and simple poles in the amplitude.

When caustics occur, classical paths
spreading out from an initial point are focused into a
unique conjugate point; in other words,
only those paths connecting the conjugate points 
are allowed classically.  
The constant $k(\lambda)$,
which is the stretching factor between
the two conjugate points, 
is found to be a useful pointer, 
along with the Morse index $m(\lambda)$, for
characterizing the intrinsic features of caustics.
Correspondingly, in quantum mechanics
the transition amplitude is nonvanishing only 
for the conjugate points (for which it diverges). 
We derived the
path-integral kernel for the amplitude in a closed form, 
expressed solely in terms of
the stretching factor, the Morse
index, and the action of (any of the) solution paths.

Contrary to the situation on caustics, once one goes
away from caustics the
classical and quantum situations become different. 
Our study focused on how the typical feature
of caustics, {\it i.e.}, the concentration of intensity
at the conjugate point, can be affected by quantum effect. 
The Gaussian slit (gedanken-)experiment using a Gaussian
wave packet shows that, although the variance itself 
is enhanced at the quantum level, 
the susceptibility of the
variance of the wave packet against initial momentum fluctuations
is suppressed.
High intensity is realized 
near caustics of two different types, the first being the 
caustics associated with the period $[0, T]$ while
the second being those with $[-\tau, T]$ which is the
entire period of our experiment.
As a quantum analogue of 
the Jacobi field, we introduced the 
susceptibility which reduces
to the ordinary Jacobi field in the classical limit
$\hbar \rightarrow 0$.  The susceptibility gives a 
measure of stability 
for the concentration, and  
our result shows that the
intensity of the amplitude is stabilized by 
quantum effect.
For practical purposes, however, we need to extend 
our analysis further to
the effects of higher order terms in the Lagrangian 
near caustics and thereby 
supplement earlier works carried out on caustics. 
Another possible direction would be to consider
these problems in higher dimensions allowing for
settings more realistic for physical application.

The classical and quantum features of caustics 
studied here have
then been employed to investigate in detail 
the role of instantons in quantum mechanics. 
Viewed as a constrained system on the plane $\R^2$,
the quantum mechanical rotor over the circle $S^1$
becomes two harmonic oscillators
sharing the same, time-dependent harmonic potential $\lambda$
given by the Lagrange multiplier field.  
Singularities in the amplitude
arise at those $\lambda$ under which caustics occur.
We however learned that 
generically no simple pole can arise, simply because 
the two stretching factors of the harmonic oscillators 
do not coincide for generic boundary conditions on $S^1$,
except when the two endpoints are the same 
or diametrically opposite with each other.  But even then, 
the zero-instanton contribution is missing in the simple poles 
to complete the correspondence (\hypo).
Even if we are confined to the former specific cases, 
and ignoring the missing piece, 
we come across the problem that it is impossible to
replace the path-integration over the potentials $\lambda$ 
by an ordinary integration over the
constant (zero-mode) part $\lambda_0$ as
originally assumed by Jevicki.  We observed that,
in the non-trivial example we looked at in section 4,
the assumption of the non-zero mode independence 
of the phase factor does not hold in general, although 
there are cases where it holds 
thanks to a certain cancellation mechanism.  From our result 
we find it hard to sustain the hypothesis even 
in the toy model, 
let alone in more general systems such as the $\Complex P^N$ model
of the original concern.  
We, however, mention that there remains a
possibility that, even though it is not strictly true, 
for some reason the assumption may still lead to a correct
answer, validating the hypothesis in the end.  We feel that
this possibility deserves a fuller study in order 
to settle down 
the issue which is important in exploring non-perturbative 
methods in quantum field theory.

\bigskip

\noindent
{\bf Acknowledgements:}
H. M. is grateful to S.~Izumiya for helpful 
discussions about caustics.
This work is supported in part by the Grant-in-Aid for Scientific
Research from the Ministry of Education, Science and Culture
(No.09740199).

\ve

\secno=0 \appno=1 \meqno=1


\noindent{\bf Appendix}
\medskip

In this Appendix we give a direct argument to show 
that, for non-critical $\lambda$, 
the phase (\genphase) of the
kernel (\genkernel) is independent of the choice of
the function $\rho(t)$.   
More explicitly, we shall show that 
the phase is given by the action,
$$
\Phi(b, a; \lambda) = I_{\rm cl}(b, T; a, 0),
\eqn\statement
$$
which is evaluated for the classical solution $x_{\rm cl}$ 
under a non-critical $\lambda$ 
obeying the boundary condition,
$x_{\rm cl}(0) = a$ and $x_{\rm cl}(T) = b$.
Prior to our argument, we recall that the function
$\rho(t)$ which satisfies 
$\rho(0) = 0$ and $\rho(T) = b - c$
is introduced to
compensate the gap 
at $t = T$ between the given endpoint $b$
and the endpoint $c$ for which 
an actual classical solution $\bar x_{\rm cl}$ exists.
Such a gap is inevitable if 
$\lambda$ is critical and $k(\lambda)\,a + s(T) \ne b$. 
However, we stress that even for non-critical $\lambda$
it is perfectly legitimate to use 
the classical solution $\bar x_{\rm cl}$ with 
$\bar x_{\rm cl}(T) = c \ne b$ for
computing the transition amplitude.
Thus, for non-critical $\lambda$, it must be that
the apparent $\rho(t)$-dependence 
in the result (\genphase) disappear and the total
phase be given by the classical action for the solution
$x_{\rm cl}(T) = b$, {\it i.e.}, (\statement).

To confirm this, we first consider  
a solution $u(t)$ of
the homogeneous equation (\motion) with $u(0) = 0$.
We may expand it in terms of the orthonormal 
eigenfunctions $\{ u_n \}$ in (\eigen) as
$$
u(t) = \sum_n c_n\, u_n(t) + u(T) \, \theta(t - T)\ .
\eqn\exu
$$
In (\exu) we introduced the step function
$\theta(t - T)$, defined by
$\theta(x) = 0$ for $x < 0$ and 
$\theta(x) = 1$ for $x \ge 0$,
to account for the condition $u(T) \ne 0$.
Differentiating (\exu) with respect to $t$ and 
setting $t = T$, we find
$$
\dot u(T) = \sum c_n\, \dot u_n(T)+  u(T)\,\delta(0) \ .
\eqn\differ
$$
Note that 
$$
c_n = \int_0^T dt\, u(t)\, u_n(t) 
 = - {1\over{E_n}} \int_0^T dt\, u(t)
   \left[{{d^2}\over{dt^2}} + \lambda(t) \right] u_n(t) 
 = -{{u(T)\, \dot u_n(T)}\over{E_n}}\ ,
\eqn\apxcoef
$$
where we used integration by parts and the equation
of motion for $u$.
Combining (\apxcoef) with (\differ) we acquire 
the formal identity,
$$
{{\dot u(T)}\over{u(T)}} = \delta(0) - 
\sum_n {{{\dot u}_n^2(T)}\over{E_n}}\ .
\eqn\apxident
$$

It is true that 
the r.h.s.~of this identity is 
ill-defined because of $\delta(0)$ and the sum
of the infinite series, but it may be made sensible by
demanding that it be valid for $\lambda = 0$.
Indeed, since we have $u(t) = \alpha t$ 
with some constant $\alpha$
for $\lambda = 0$, the demand suggests
$$
1 = T \, \delta(0) - 2 \sum_n 1\ .
\eqn\interpret
$$
This allows us to \lq renormalize' (\apxident)
to get
$$
{{\dot u(T)}\over{u(T)}} = {1 \over T} 
- \sum_n \left( 
{{{\dot u}_n^2(T)}\over{E_n}} - {2 \over T} \right) \ .
\eqn\renident
$$
This time the r.h.s.~is well-defined, since 
each of the terms in the infinite series becomes small
sufficiently fast when $n$ becomes large, as seen from
the asymptotic behaviour of the
eigenfunctions [\CH, \MF].  

Consider next the expansion of the function
$\rho(t)$ appearing in (\genphase) analogous to
(\exu),
$$
\rho(t) = \sum_n d_n\, u_n(t) + \rho(T) \, \theta(t - T)\ ,
\eqn\exrho
$$
where $d_n = \int_0^T dt\, \rho(t)\, u_n(t)$. 
Substituting (\exrho) in the phase (\genphase) we find
$$
\Phi(b, a; \lambda) = I[ \bar{x}_{\rm cl} ] 
+ \dot{\bar{x}}_{\rm cl}(T) \rho(T) 
+ {1\over 2} \rho^2(T) 
\left\{ 
\delta(0) - \sum_n {{{\dot u}_n^2(T)}\over{E_n}}
\right\}\ .
\eqn\phident
$$
Now, using the identity (\apxident), and 
noticing that the solution $u$ may be given by
$u = \beta(x_{\rm cl} - \bar{x}_{\rm cl})$ with  
$\beta$ a constant, we arrive at
$$
\eqalign{
\Phi(b, a; \lambda) 
&= \frac12 \bar{x}_{\rm cl} \dot{\bar{x}}_{\rm cl}\Big\vert^T_0
   + \frac12 \rho(T) 
   \bigl\{ \dot x_{\rm cl}(T) 
   + \dot{\bar{x}}_{\rm cl}(T)\bigr\} \cr  
&= I_{\rm cl}(b, T; a, 0) + \frac12  
   \bigl\{ a\, \dot x_{\rm cl}(0) - a\, \dot{\bar{x}}_{\rm cl}(0)
   + b\, \dot{\bar x}_{\rm cl}(T) 
    - c\, \dot{x}_{\rm cl}(T) \bigr\} \ .
}
\eqn\semifinal
$$
The relation (\statement) then follows by use of Lemma 1.

\ve

\bsk

\vfill\eject

  \vfill\eject\immediate\closeout\reffile
  \centerline{{\bf References}}\bigskip\frenchspacing%
  \input refs.tmp\vfill\eject\nonfrenchspacing

\bye